\documentclass[twocolumn, times]{aastex631}

\usepackage{soul}

\usepackage{CJK}

\newcommand       \cm           {\,{\rm cm}}

\newcommand       \erg          {\,{\rm erg}}

\newcommand       \g            {\,{\rm g}}
\newcommand       \K            {\,{\rm K}}
\newcommand       \km           {\,{\rm km}}

\newcommand       \Mpc          {\,{\rm Mpc}}
\newcommand       \s            {\,{\rm s}}

\newcommand       \yr       {\,{\rm yr}}

\newcommand       \mum          {\,{\rm \mu m}}

\newcommand       \Msun         {\,{M_\odot}}

\newcommand       \Lsun         {\,{L_\odot}}

\newcommand       \LIR          {L_{\rm IR}}

\newcommand       \simali       {\sim\,}

\newcommand       \NCaro      {N_{\rm C,aro}}
\newcommand       \NCali      {N_{\rm C,ali}}
\newcommand       \alifrac      {\eta_{\rm ali}}
\newcommand{\cosmology}{$\Omega_m = 0.287$ and $H_0=69.3\km\s^{-1}\Mpc^{-1}$}

\received{February 20, 2025}
\revised{April 14, 2025}
\accepted{May 5, 2025}

\submitjournal{ApJ}

\begin{document}

\title{\large Unveiling the Aromatic and Aliphatic Universe
         at Redshifts $z$\,$\simali$0.2--0.5 with JWST NIRCam/WFSS
         }

\author[0000-0002-6221-1829]{Jianwei Lyu (\begin{CJK}{UTF8}{gbsn}吕建伟\end{CJK})}
\affiliation{Steward Observatory, University of Arizona,
933 North Cherry Avenue, Tucson, AZ 85721, USA;
{\sf jianwei@arizona.edu}}

\author[0000-0002-6605-6512]{Xuejuan Yang}
\affiliation{Department of Physics,
    Xiangtan University,
411105 Xiangtan, Hunan Province, China;
{\sf xjyang@xtu.edu.cn}}

\author[0000-0002-1119-642X]{Aigen Li}
\affiliation{Department of Physics and Astronomy,
                  University of Missouri,
              Columbia, MO 65211, USA;
                  {\sf lia@missouri.edu}}

\author[0000-0002-4622-6617]{Fengwu Sun}
\affiliation{Center for Astrophysics $|$ Harvard \& Smithsonian, 60 Garden St., Cambridge MA 02138 USA}

\author[0000-0003-2303-6519]{George H. Rieke}
\affiliation{Steward Observatory, University of Arizona,
933 North Cherry Avenue, Tucson, AZ 85721, USA}

\author[0000-0002-8909-8782]{Stacey Alberts}
\affiliation{Steward Observatory, University of Arizona,
933 North Cherry Avenue, Tucson, AZ 85721, USA}

\author[0000-0003-4702-7561]{Irene Shivaei}
\affiliation{Centro de Astrobiolog\'ia (CAB), CSIC-INTA, Ctra. de Ajalvir km 4, Torrej\'on de Ardoz, E-28850, Madrid, Spain}


%
%

\begin{abstract}
Utilizing deep NIRCam/WFSS data from JWST's FRESCO program, we spectroscopically survey the 3.3$\mum$ aromatic and 3.4$\mum$ aliphatic C--H stretching emission bands of polycyclic aromatic hydrocarbon (PAH) molecules in galaxies at redshifts $z$\,$\simali$0.2--0.5. Unlike pre-JWST studies, largely limited to infrared 
(IR)-bright galaxies ($L_{\rm IR}\gtrsim10^{11}\Lsun$) at $z\lesssim0.1$, 
we probe 200 galaxies down to $L_{\rm IR}$\,$\simali$$10^{8.5}$--$10^{10}\Lsun$ well 
beyond the local Universe. The 3.3$\mum$ emission is detected at $\geq$3-$\sigma$ in 88 out of 187 galaxies, correlating 
tightly with galaxy IR luminosity and star formation rate (SFR) and confirming the 3.3$\mum$ PAH as a viable SFR tracer. Despite a large scatter, the 3.3$\mum$-to-IR luminosity ratio ($L_{3.3}/\LIR$) exhibits a strong metallicity dependence with a drop of $L_{3.3}/\LIR$ by a factor of $\gtrsim10$ at 12+log(O/H)$\sim$8.4--8.5 towards lower metallicities. 
The 3.4$\mum$ emission is detected in 37 out of 159 galaxies, with the 3.4$\mum$-to-3.3$\mum$ luminosity ratio
($L_{3.4}/L_{3.3}$) spanning from $\simali$0.05 to $\simali$0.58 (median 
$\simali$0.19), corresponding to PAH aliphatic fractions of $\simali$0.78\%--8.3\% (median $\simali$2.9\%) in terms of fractional carbon atoms in aliphatic units. 
While $L_{3.4}/L_{3.3}$ does not depend significantly on 
redshift, stellar mass, metallicity, or galaxy morphology, it does decrease with various SFR tracers, suggesting that
ultraviolet photons in active star-forming regions may strip aliphatic sidegroups from PAH molecules. Our study showcases the unique power of JWST's
NIRCam/WFSS to systematically map PAH aromatic and aliphatic content in
statistically significant, less-biased galaxy samples, providing critical insights 
into PAH chemistry and its connection to galaxy properties. 
\end{abstract} 


\section{Introduction}\label{sec:intro}
Polycyclic aromatic hydrocarbon (PAH) molecules 
are ubiquitous and widespread in the Universe, 
serving as key tracers of star formation and galaxy evolution. 
These organic molecules emit distinctive infrared (IR) 
features at wavelengths of 3.3, 6.2, 7.7, 8.6, 11.3 and
12.7$\mum$, collectively also known as
the ``unidentified infrared emission'' (UIE) bands
\citep[see][]{Tielens2008, Li2020}.
These UIE bands are now commonly believed to arise
from the stretching and bending vibrational modes
of PAH molecules, which consist of aromatic carbon
and hydrogen atoms arranged in benzene ring structures
\citep{Leger1984, Allamandola1985}.
The intensities of these emission bands are tightly
correlated with star formation rate (SFR), allowing 
one to probe star formation activity even in highly
obscured environments \citep[e.g.,][]{Pope2008, Shipley2016, Xie2019}.
Furthermore, PAHs dominate the heating of the interstellar gas
by providing photoelectrons and influence the thermal balance
and chemistry of the surrounding interstellar medium
\citep[ISM;][]{Tielens2008, Draine2007}.
Understanding the nature, origin and evolution of PAHs
across cosmic time and galaxy properties
offers crucial insights into the interplay
between galaxy evolution, dust, and gas.

The 3.3$\mum$ PAH emission band,
originating from the aromatic C--H stretches,
serves as a critical diagnostic of small, neutral PAHs
containing approximately 20--30 carbon atoms
\citep[see][]{Draine2007, Draine2021}.
%
This emission band is sometimes accompanied
by a weak satellite emission feature at 3.4$\mum$,
which is generally thought to arise
from the aliphatic C--H stretch.
This implies that astronomical PAHs often include
an aliphatic component, e.g., aliphatic sidegroups
like methyl (--CH$_3$) attached as functional groups
to PAHs \citep{Yang2017, Allamandola2021}.
Alternative assignment of the 3.4$\mum$ feature
includes anharmonicity \citep{Barker1987}
or superhydrogenation
\citep{Bernstein1996, Sandford2013, Yang2020}.
Despite its potential to reveal critical insights
into PAH structure and evolution,
the 3.4$\mum$ feature is significantly weaker
than the 3.3$\mum$ feature, complicating its detection.
Previous extragalactic observations of the 3.4$\mum$ band
have primarily been limited to local IR-luminous galaxies
and often required high signal-to-noise spectra to confirm
its presence.
{ 
In the era of the {\it Infrared Space Observatory} (ISO), the 3.3 and 3.4$\mum$ features were detected with its {\it Short Wavelength Spectrometer} (SWS) in several
nearby, bright galaxies, including
M82, a prototypical starburst galaxy
with a strong superwind halo;
NGC 253, a barred spiral galaxy with a high
level of circumnuclear starburst activity;
the 30 Dor region, a large, massive and luminous H{\sc ii} region in the Large Magellanic Cloud; and Circinus, a nearby spiral Seyfert 2 galaxy \citep{Sturm2000}. In normal galaxies, only a weak emission continuum in the 2.5--4.9$\mum$ wavelength range was seen in the spectra obtained with ISO's Photopolarimeter (ISOPHOT; \citealt{Helou2000, Lu2003}). With an improved sensitivity, the Japanese AKARI mission appreciably enlarged the sample of galaxies for which the 3.3 and 3.4$\mum$ features were detected \citep[see e.g.,][]{Imanishi2010, Kondo2012, Yamagishi2012, Lee2012, Inami2018, Lai2020}. We note that, although {\it Spitzer's Infrared Spectrograph} (IRS) did observe the 3.3 $\mum$ PAH features in a few IR bright galaxies at $z\sim$2--3  \citep{Sajina2009, Siana2009}, unfortunately, the 3.4 $\mu$m C--H stretching features are too weak to be detected.
}

With the successful launch and operation of JWST,
the study of the 3.3 and 3.4$\mum$ aromatic
and aliphatic C--H stretches of PAHs
in extragalactic sources has entered a new era.
For example, the high-spatial-resolution and
sensitive narrow/medium band NIRCam images allow
the construction of detailed 3.3$\mum$ PAH
emission maps of low-$z$ galaxies up to $z\sim0.5$
\citep[e.g.,][]{Sandstrom2023, Gregg2024, Alberts2024}.
NIRSpec observations enable the study
of the 3.3 and 3.4$\mum$ emission bands
along with other molecular and atomic lines
well extended into the galactic halo in M82
in great detail (J.W.~Lyu et al.\ in preparation).
With the sensitive mid-IR spectral observing
capability of MIRI/MRS, the detection of
the 3.3$\mum$ PAH emission has been even
reported in a gravitationally-lensed galaxy at
redshift $z\approx4.2$ \citep{Spilker2023},
when the Universe was only $\simali$11\%
of its current age. Based on a MIRI imaging survey,
\citet{Shivaei2024} conducted a systematic study
of the evolution of PAH strength from the spectral
energy distribution (SED) analysis among galaxies
up to cosmic noon, further demonstrating
the power of JWST in statistical studies of PAHs.

In this work, we will demonstrate that the Wide-Field
Slitless Spectroscopy (WFSS) mode of JWST/NIRCam
has also opened new avenues for PAH studies.
Unlike targeted observations, WFSS enables blind
surveys across large areas of the sky, which capture
spectra of all sources within the field of view,
offering the opportunity to characterize the PAH behavior
in statistically significant and unbiased samples of galaxies.
Thanks to the suprior sensitivity of NIRCam,
with a modest exposure time ($\simali$1\,hr),
this allows for the detection of PAH emission in
a previously unexplored population of faint galaxies
with IR luminosity $L_{\rm IR}$ as low as 
$10^{8.5}$--$10^{10}\Lsun$ at $z\lesssim0.5$, 
greatly surpassing the limits of earlier 
AKARI observations with $L_{\rm IR}>10^{12}\Lsun$
at similar redshifts \citep[e.g.,][]{Imanishi2010, Ichikawa2014}. 
By covering diverse galaxy populations, 
including those with very low star formation rates, 
JWST/NIRCam WFSS broadens the scope of PAH studies, 
providing a more representative picture of 
the galaxy population and facilitating 
comprehensive statistical analyses.

Based on NIRCam/WFSS F444W observations of
two GOODS/CANDELS fields, we report here detections
of the 3.3 and 3.4$\mum$ emission bands
in galaxies at $z$\,$\simali$0.2--0.5
from the JWST Cycle 1 legacy program FRESCO
(First Reionization Epoch Spectroscpically
Complete Observations; \citealt{Oesch2023}).
Although FRESCO was designed for the study
of very distant Universe, its NIRCam grism observations
provide high-quality spectra at 3.8--5$\mum$ for objects
in the survey footprint, enabling the detection of
the 3.3$\mum$ aromatic and weaker 3.4$\mum$
aliphatic C--H stretches, if present, in any galaxies
at $z$\,$\simali$0.2--0.5 within its survey footprint.
In addition, the two fields covered by FRESCO have
the most extensive observations across the whole
electromagnetic spectrum with both ground-based
and space-based facilities over the past two decades,
allowing accurate characterizations of various galaxy
properties, e.g., stellar population, 
active galactic nuclei (AGN) content,
galaxy morphology, star formation activities
in a consistent manner. Thanks to all these,
we are able to conduct a comprehensive study
on the behaviors of the 3.3 and 3.4$\mum$
emission bands among galaxies with a wide range
of properties based on a statistical sample.

We structure this paper as follows.
In \S2 we briefly describe the data
and the sample selection. \S3 describes
how we conduct various measurements
such as the strengths of the 3.3 and 3.4$\mum$
emission bands and put constraints
on the galaxy properties. In \S4 we explore
the behaviors of the 3.3 and 3.4$\mum$ bands
and derive the aromatic and aliphatic contents
of PAHs and discuss how they vary among
galaxy properties (e.g., star formation rates,
metallicities, and morphologies).
Finally, a summary is given in \S5.

Throughout this work, we assume
a flat $\Lambda$CDM cosmology
with \cosmology \citep{WMAP9}.

\section{Data and Sample Selection}\label{sec:data}
\subsection{The FRESCO Survey and
                    NIRCam/WFSS Data
                    Reduction}\label{sec:fresco-reduction}
FRESCO is a 53.8\,hr medium-sized JWST Cycle 1 legacy program that targets at
both the GOODS-S and GOODS-N fields with deep NIRCam/grim WFSS observations. For each field,
the NIRCam WFSS observations were conducted to cover a footprint of 62
arcmin$^2$ in the F444W filter that provides $R\sim1600$ slitless spectra at
$\lambda$\,$\simali$3.8--5.0$\mum$ for sources within the survey
footprint. With an exposure time of $\simali$2\,hr per pointing,
the reached 5-$\sigma$ emission-line sensitivity for compact sources is
$\simali$$2\times10^{-18}\erg\s^{-1}\cm^{-2}$. In addition to the F444W grism
data, NIRCam direct imaging observations in F182M, F210M and F444W with a
similar footprint were also obtained. We refer to \citet{Oesch2023} for any
additional details about the design and configuration of this program.

We adopted the publicly available NIRCam WFSS data processing routine presented
in \citet{Sun2023}\footnote{%
  \href{https://github.com/fengwusun/nircam_grism}{https://github.com/fengwusun/nircam\_grism}
}. Here we only provide a brief summary. In the first round, the data were
processed with the standard JWST stage-1 calibration pipeline v1.11.2. For each
individual exposure, we assigned world coordinate system (WCS) information,
performed flat-fielding, and removed the $\sigma$-clipped median sky background.

Given that our sources are reasonably bright and extended, we did not perform
any additional background subtraction during the spectral extraction. We
optimally extracted 1-D spectra of our targets using their surface brightness
profile in the F444W band \citep{Horne1986}. Since the data are quite deep,
contamination from other sources can be seen in many final extracted 2-D and 1-D
spectra as evident from the abnormal continuum shape. A detailed contaminant
subtraction is very complicated as it requires modeling of the 2-D spectra of
$\simali$10$^4$ sources across the field, which is beyond the scope of the
current paper. As a result, our spectral analysis later will be limited to the
sources and wavelength ranges where the contamination is not obvious.  


\subsection{Sample Selection and Data Collection}
In the pre-JWST era, the two GOODS fields have been extensively observed with
multi-band images from the ultraviolet (UV) to the mid-IR as well as optical to
near-IR spectroscopic follow-ups from both ground-based and space-based missions
that allow comprehensive characterizations of the source properties. To
construct the parent sample for the FRESCO PAH identifications, we adopted the
corresponding 3D-HST catalogs in these fields \citep{Skelton2014}, which
collected the HST/grism redshifts, ground-based spectroscopic redshifts and
photometric redshifts. We required the galaxies to have redshifts at
$z$\,$\simali$0.2--0.45 for the spectral coverage of the 3.3$\mum$ PAH feature
in the NIRCAM/WFSS F444W filter ($\simali$3.9--5.0$\mum$). Within the FRESCO
survey footprint, we ended up 200 galaxies with NIRCam/WFSS grism spectra in total with 101 in GOODS-S and 99
in GOODS-N at this redshift range. 

In addition to the FRESCO spectral extraction
as described in \S\ref{sec:fresco-reduction},
we collected various ancillary data for these galaxies
to characterize their properties. Besides the ground-based,
HST and Spitzer/IRAC photometry
at $\lambda$\,$\simali$0.3--8.0$\mum$
provided by the 3D-HST catalogs,
we also cross-matched the sample with available
Spitzer/IRS 16$\mum$ and Spitzer/MIPS 24$\mum$
photometry measurements at longer wavelengths
following \citet{Lyu2022} to enable the estimation of galaxy IR luminosity and
to improve the identification of obscured AGNs. To estimate the possible AGN
contribution in these galaxies, we have matched the sample against the bright AGN
catalogs in GOODS-S \citep{Lyu2022} and GOODS-N (Lyu et al., in prep), which are
based on a comprehensive search from the X-ray to the radio bands with the
deepest data accessible. Lastly, to characterize the galaxy morphology, we
collected the JWST/NIRCam F444W images obtained by the FRESCO program. These
images have a PSF FWHM of $0.145\arcsec$ corresponding to 0.4--0.8\,kpc, and
cover the rest-frame 3--3.7$\mum$ for galaxies at $z$\,$\simali$0.2--0.45,
allowing a good characterization of most galaxy morphology structures with
little dust obscuration.

\section{Measurements of Galaxy Properties and PAH Spectral Properties}
\subsection{Galaxy Properties}
The UV to mid-IR SEDs can be used to decipher many key galaxy properties such as
star formation history, stellar age, star formation rate, extinction, and
possibly as well AGN content. For the SED analysis, we adopted a modified
version of {\sf Prospector} \citep{Johnson2021} that includes the semi-empirical
models of AGN component and galaxy dust emission as detailed in \citet{Lyu2024}.
For the stellar component, we assumed the Kroupa initial mass function and a
delayed-$\tau$ star formation history with the \citet{Kriek2013} dust
attenuation law. For the galaxy dust emission, we adopted the $\log(L_{\rm
IR}/L_\odot)=11.25$ template of \citet{Rieke2009} without the requirement of
energy balance. Regarding the AGN component, we used the AGN empirical model
with a hybrid extinction law and various narrow and broad emission lines as described in \citep{Lyu2024}. Readers
are highly recommended to check the relevant sections in \citet{Lyu2024} to
learn more details about these configurations and justifications. From the SED
fittings, we can get useful constraints on the stellar mass ($M_\star$), stellar
age ($t_{\rm age}$), $e$-folding time for the star formation history
($\tau_{\star}$), dust attenuation level of the stellar continuum
($A_{V,\star}$), and galaxy IR luminosity ($L_{\rm IR, gal}$). If the AGN
contribution is strong with enough SED coverage, AGN properties such as
bolometric luminosity ($L_{\rm AGN}$) and obscuration levels ($\tau_{\rm IR}$,
$\tau_{\rm OPT}$) can be also reasonably constrained. In contrast to the original {\sf Prospector} code, we do not introduce energy balance in our fittings.

Due to the complicated nature of AGN and the limitations of the photometry data,
SED fittings will not reveal all AGNs in the sample. Luckily, a comprehensive
multi-wavelength AGN search in these fields has been carried out by
\citet{Lyu2022, Lyu2024} in GOODS-S and by Lyu et al. (in preparation) in
GOODS-N. To improve the AGN census, these works utilized the deepest X-ray,
optical, mid-IR and radio data available and employed eight selection methods
that consider the source X-ray luminosity, X-ray to radio flux ratio, mid-IR
color, UV-to-mid-IR SED, radio loudness, radio slope, X-ray/optical variability,
spectral emission line properties. We cross-matched our sample of galaxies with
these work, and found 24 AGN candidates in GOODS-S and 11 AGN candidates in
GOODS-N. However, most of these AGNs are identified only through the
radio--to--X-ray ratio, indicating the AGN is not strong and unlikely
influence the PAH behaviors. Eight objects are identified as
SED AGNs (as indicated in Table~\ref{tab:sample}), however, their AGN component is obscured at 
shorter wavelength. None
of our sample have been classified as AGN in X-ray. In other words, the AGN
contribution to the UV-to-near-IR SED is quite low, so the SED analysis of the
galaxy stellar properties are not strongly influenced. Given the lack of strong AGNs in 
the sample, we will not differentiate objects with and without an AGN.

Although we allow stellar metallicity to vary in our {\sf Prospector} fits, due
to its strong degeneracy with stellar population and dust attenuation, the
fitted values are not very useful. Direct measurements of gas-phase metallicity
requires high-quality rest-frame optical spectra that cover a wide range of
wavelength, which are not available for the majority of our sources. Following
the strategy described in \citet{Shivaei2024}, we estimated metallicities,
12+log(O/H), with the fundamental metallicity relation (FMR) between stellar
mass, SFR, and metallicity as presented in \citet{Sanders2021}. For 13 out of
102 GOODS-N galaxies in our sample, we found measurements of their gas-phase
metallicity from \citet{Kobulnicky2004} based on spectral constraints, and these
measurements are consistent with the values estimated from FMR within
measurement uncertainties.

There are two ways to estimate the galaxy SFRs. Firstly, we can derive the SFR
based on the stellar population synthesis results from the {\sf Prospector} SED
fittings, assuming a timescale of 10 Myr. The results have been found to be in
good agreement with popular SFR tracers such as H$\alpha$ with a scatter
$\lesssim$0.1--0.2 dex \citep{Leja2017}. Secondly, we can estimate the SFR based
on the galaxy IR luminosity following the empirical SFR calibration of
\citet{Kennicutt1998}. For objects without evidence of AGN from SED analysis, we matched the MIPS 24$\mu$m photometry by
the $\log(L_{\rm IR}/L_\odot)=11.25$ IR galaxy template of \citet{Rieke2009}
which has been found to be the best-match of galaxies at higher redshifts
\citep{Rujopakarn2013}, and compute the total IR luminosity at 8--1000$\mum$.\footnote{90 galaxies in our sample have {\it Herschel} far-IR detections \citep{Elbaz2011}. However, most of them (69 out of 90) have nearby bright sources that would contaminate the flux measurements, so we decide to use the template-based approach, rather than conducting a complete IR SED fitting throughout the sample.}
For galaxies without MIPS 24$\mum$ detection or with an AGN component, we relied on the galaxy IR
luminosity derived from the SED fittings where the stellar or AGN contribution
at shorter wavelengths is modeled and subtracted. For 91 galaxies, the SED fitting did not reveal any notable galaxy dust emission and we derived an upper limit of the galaxy IR luminosity by scaling the dust template to the MIPS 24$\mum$ upper limit. We have compared the IR
luminosity values and the SED-derived SFRs and found that they follow closely to
the empirical relation described in \citet{kennicutt2012}.

We made the NIRCam/F444W image cutouts of each galaxy and conducted visual
classifications for the galaxy morphology. Following the strategies described in
\citet{Kocevski2015}, we group the galaxies into six morphology categories: ud
-- undisturbed disk; dd -- disturbed disk; sp -- spheroid; ps -- point source;
ir -- iregular;  me -- merger.

In Table~\ref{tab:sample}, we summarize the various galaxy properties of the
sample together with the source ID, redshifts and coordinates in the sky. 

\begin{deluxetable*}{cccccccccccc}
\tablecaption{Properties of Galaxies at $z$\,=\,0.2--0.5 within FRESCO footprint}
\tablehead{
\colhead{ID} &
\colhead{RA} &
\colhead{Dec} &
\colhead{$z$} &
\colhead{$\log(M_\star/M_\odot)$} &
\colhead{$t_{\rm age}$} &
\colhead{$\tau_\star$} &
\colhead{SFR$_{\rm SED}$} &
\colhead{12+log(O/H)} &
\colhead{$\log\left(L_{\rm IR}/L_\odot\right)$} &
\colhead{morphology} &
\colhead{AGN?} \\
\colhead{} &
\colhead{deg} &
\colhead{deg} &
\colhead{} &
\colhead{} &
\colhead{Gyr} &
\colhead{Gyr} &
\colhead{$M_\odot~{\rm yr}^{-1}$} &
\colhead{} &
\colhead{} &
\colhead{} &
\colhead{} 
}
\startdata 
GS-08629 &   53.1835632 & -27.8621082 &  0.279 &  10.98 &  10.10 &   5.68 &  45.51 &  8.80  &  11.10  &  ir  & N \\ 
GS-09402 &   53.1703529 & -27.8667488 &  0.359 &   9.48 &   2.18 &   0.82 &   9.84 &  8.39  &  9.35   &  ud  & N \\ 
GS-09755 &   53.1655006 & -27.8652668 &  0.413 &   9.56 &   1.31 &   2.37 &   2.15 &  8.60  &  10.58  &  me  & N \\ 
GS-09883 &   53.1525574 & -27.8647594 &  0.347 &   9.92 &   5.64 &   1.93 &  12.28 &  8.56  &  10.46  &  ud  & N \\
     $\cdots$  &   $\cdots$   &  $\cdots$    &  $\cdots$     &     $\cdots$    &       $\cdots$        &   $\cdots$   &   $\cdots$   &   $\cdots$  
\enddata
\tablecomments{
Only a portion of this table is shown here to demonstrate its form and content. 
A machine-readable version of the full table is available online.}
\label{tab:sample}
\end{deluxetable*}

\begin{figure}
\begin{center}
\includegraphics[width=1.0\hsize]{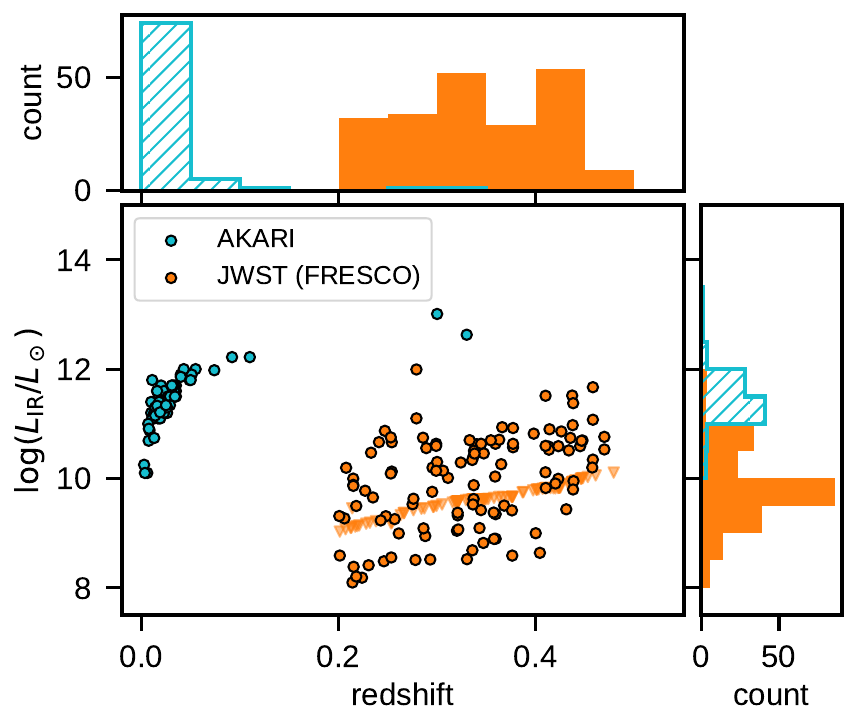}
\end{center}
\caption{Galaxy IR luminosity versus redshift for the AKARI sample (blue dots; \citealt{Imanishi2010, Ichikawa2014}) and the JWST/FERSCO sample in this work (orange dots for galaxies with $L_{\rm IR}$ measurements and orange triangles for galaxis with $L_{\rm IR}$ upper limits).   
         }
         \label{fig:lir_z_dist}
\end{figure}

To illustrate how our sample differs from previous studies
based on AKARI data, we compare in Figure~\ref{fig:lir_z_dist}
the IR luminosity versus redshift distribution
of our selected galaxy sample
with that of the AKARI sample
\citep{Imanishi2010, Ichikawa2014}.
It is clear that the previous AKARI study
was largely limited to galaxies at $z\lesssim0.1$
with very few at $z$\,$\simali$0.1--0.3.
Besides a limited number of galaxies
with $\log(L_{\rm IR}/L_\odot)$\,$\simali$10--11
at very low-z, most of the AKARI galaxies
belong to luminous IR galaxies
(LIRGs; $\log(L_{\rm IR}/L_\odot)$=11--12)
or ultraluminous IR galaxies
(ULIRGs; $\log(L_{\rm IR}/L_\odot)>12$).
In contrast, with the deep NIRCam/WFSS data
and other multi-wavelength data in these deep fields,
we are now able to explore the 3.3 and 3.4$\mum$
emission bands down to $\log(L_{\rm IR}/L_\odot)\lesssim11$
statistically at $z$\,$\simali$0.2--0.45.

\subsection{NIRCam/WFSS Spectral Analysis
                    and PAH Feature Characterization} 
%
%
%
%
%

%

For each source, depending on the wavelength coverage
of the usable spectrum, we fit the observed spectrum
in terms of one or more Drude profiles combined
with an underlying linear continuum:
\begin{equation}\label{eq:drude}
F_\lambda = a_0 + a_1 \lambda + \sum_j
\frac{P_j\,\times\,\left(2\gamma_j/\pi\right)}
{\left(\lambda-\lambda_{{\rm o},j}^2/\lambda\right)^2+\gamma_j^2} ~~,
\end{equation}
where $a_0$ and $a_1$ are the coefficients
of the linear continuum; $\lambda_{{\rm o},j}$
and $\gamma_j$ are the central wavelength
and width of the $j$-th Drude profile;
$P_j$, the power emitted from the $j$-th Drude profile
(in unit of $\erg\s^{-1}\cm^{-2}$),
is obtained by integrating the emission feature
over wavelength:
\begin{equation}\label{eq:P_j}
P_j = \int_{\lambda_j}\Delta F_\lambda\,d\lambda ~~.
\end{equation}
As Drude profiles are expected for damping harmonic oscillators, we approximate
the 3.3 and 3.4$\mum$ emission bands as Drude profiles. In some objects (e.g.,
GOODS-S\,21978), an additional weak band at $\simali$3.47$\mum$ is also present
and this band is also included as a Drude profile.
In some sources, this 3.47$\mum$ sub-feature appears as a broad plateau,
substantially broader than the typical satellite features of the 3.3$\mum$ band
at 3.40, 3.43, 3.47, 3.51, and 3.56$\mum$. This 3.47$\mum$ broad plateau appears
to correlate with the 3.3$\mum$ band \citep{Pilleri2015, Lai2020}. In contrast,
the 3.47$\mum$ band stands pronouncedly in other sources as a sharp feature,
resembling that seen in the planetary nebula IRAS\,21282\,+\,5050 (see Figure~3
of \citealt{Geballe1994}). In this case, the 3.47$\mum$ band (together with the
other weak bands at 3.43, 3.51, and 3.56$\mum$) is generally attributed to
aliphatic C--H stretch (see \citealt{Yang2017}), although the anharmonicity
hyphothesis considers all these bands as aromatic (see \citealt{Barker1987}).
%
%
Due to this complexity,
we will only count the 3.4$\mum$ emission
when we discuss the PAH aliphatic content.
Also, the strength of the 3.47$\mum$ feature 
is sometimes uncertain because of contamination 
and possible continuum variations.

Due to the nature of slitless spectra,
the profile of all these PAH emission bands
can be further broadened by the extended galaxy morphology since the PAH emission can come from different locations.
As a result, we have convolved all the fitted features
with the galaxy 1-D profile
along the dispersion direction derived from the F444W image to improve the fittings.
We note that the total flux of the band within the aperture
would not be affected by this effect
once the underlying continuum is properly subtracted.

For all the sources, we have utilized Monte-Carlo methods to produce 1000 mock
spectra by perturbing the observed spectrum with the measured 1-$\sigma$
uncertainties. We measured the PAH feature strength in each mock spectrum and
compute the standard deviations of the distribution as the measurement
uncertainties. In addition, we visually inspected both 2-D and 1-D NIRCam/WFSS
spectra and dropped sources with strong nearby source contamination or
wavelength coverage too limited for meaningful fittings. In total, we have
NIRCam/WFSS spectra of 187 galaxies and 159 galaxies that can be used to
constrain the 3.3 and 3.4$\mum$ features. 

\begin{figure*}[h!]
\centering
\includegraphics[width=0.495\hsize]{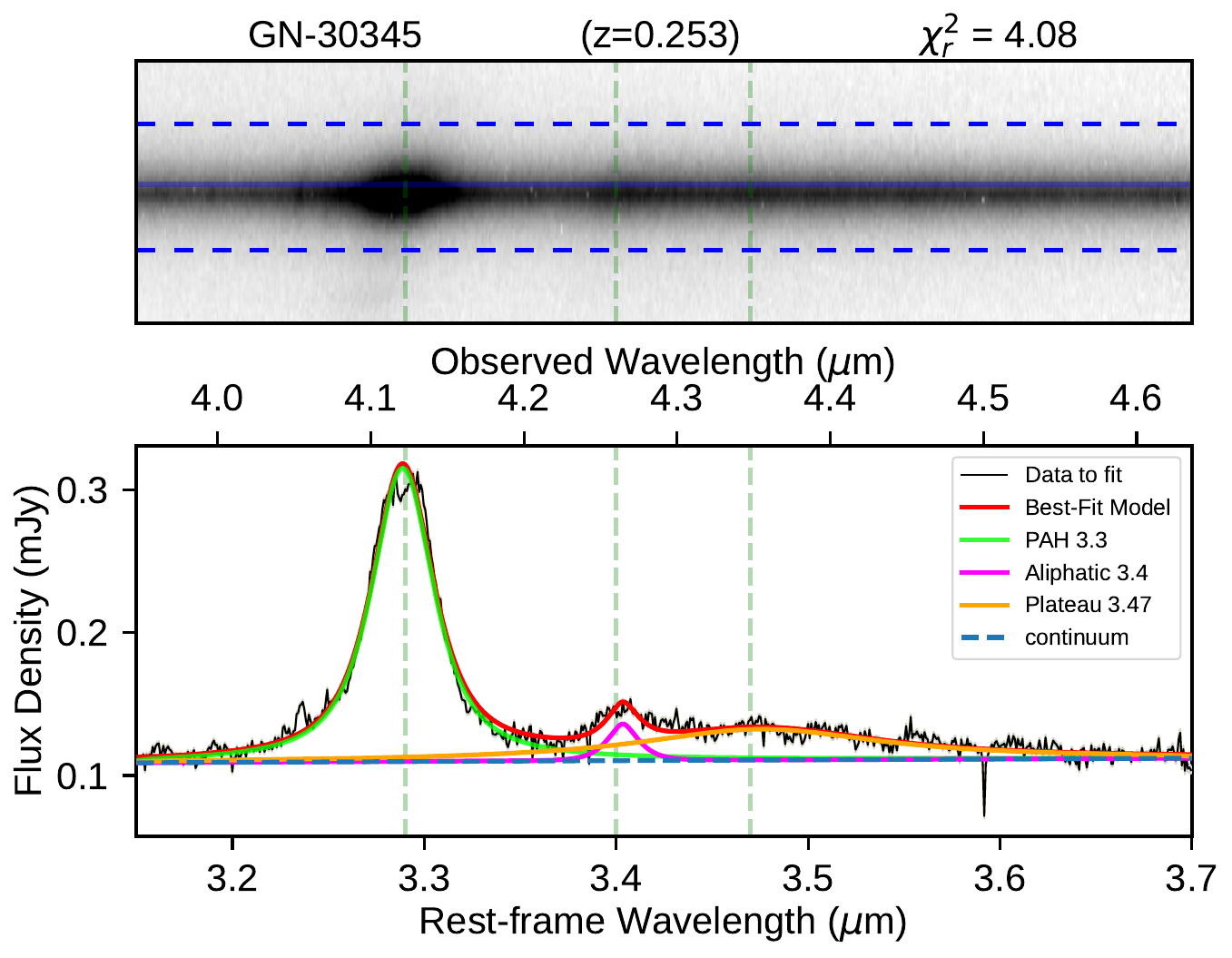}%
\includegraphics[width=0.495\hsize]{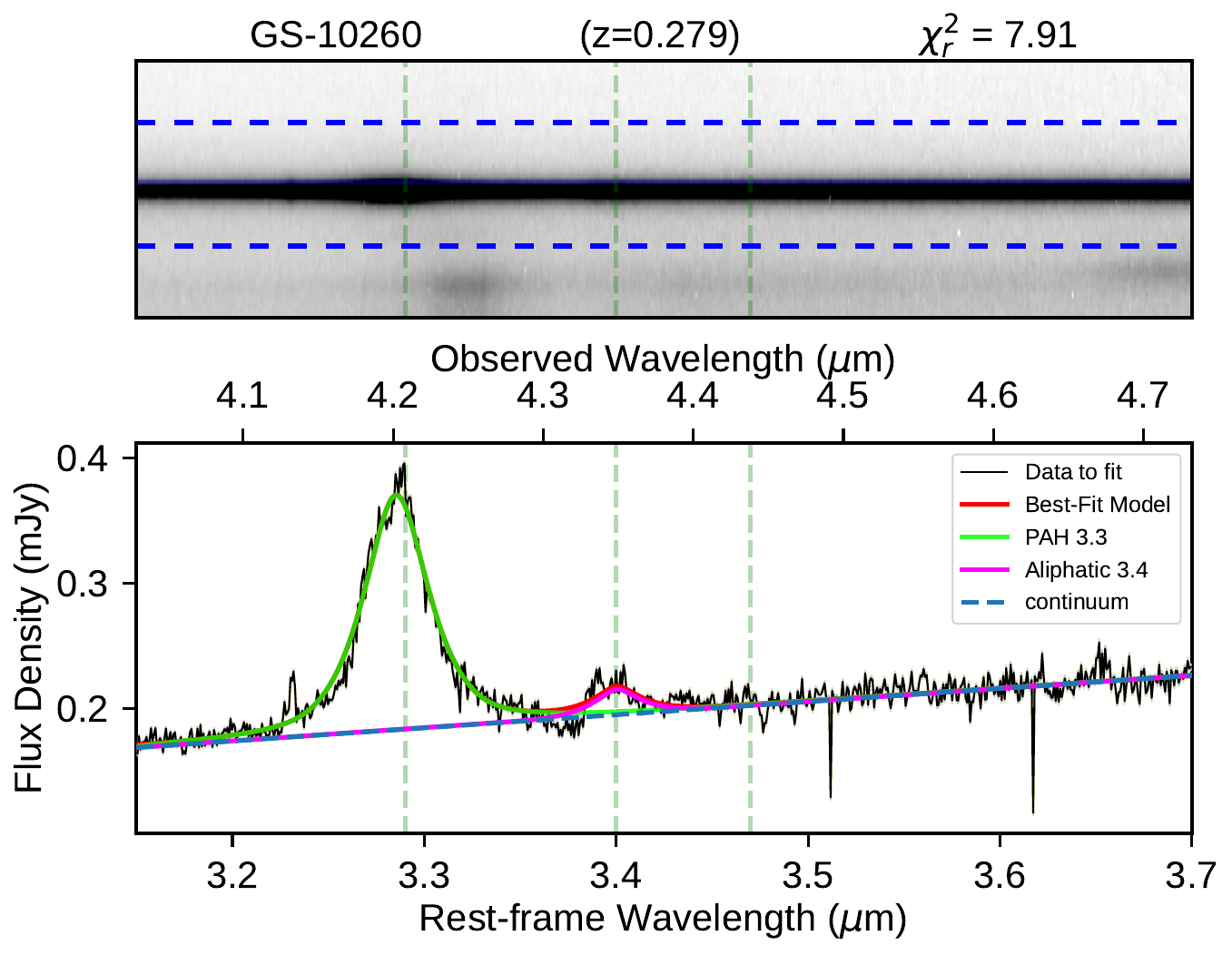}%
\\
\includegraphics[width=0.495\hsize]{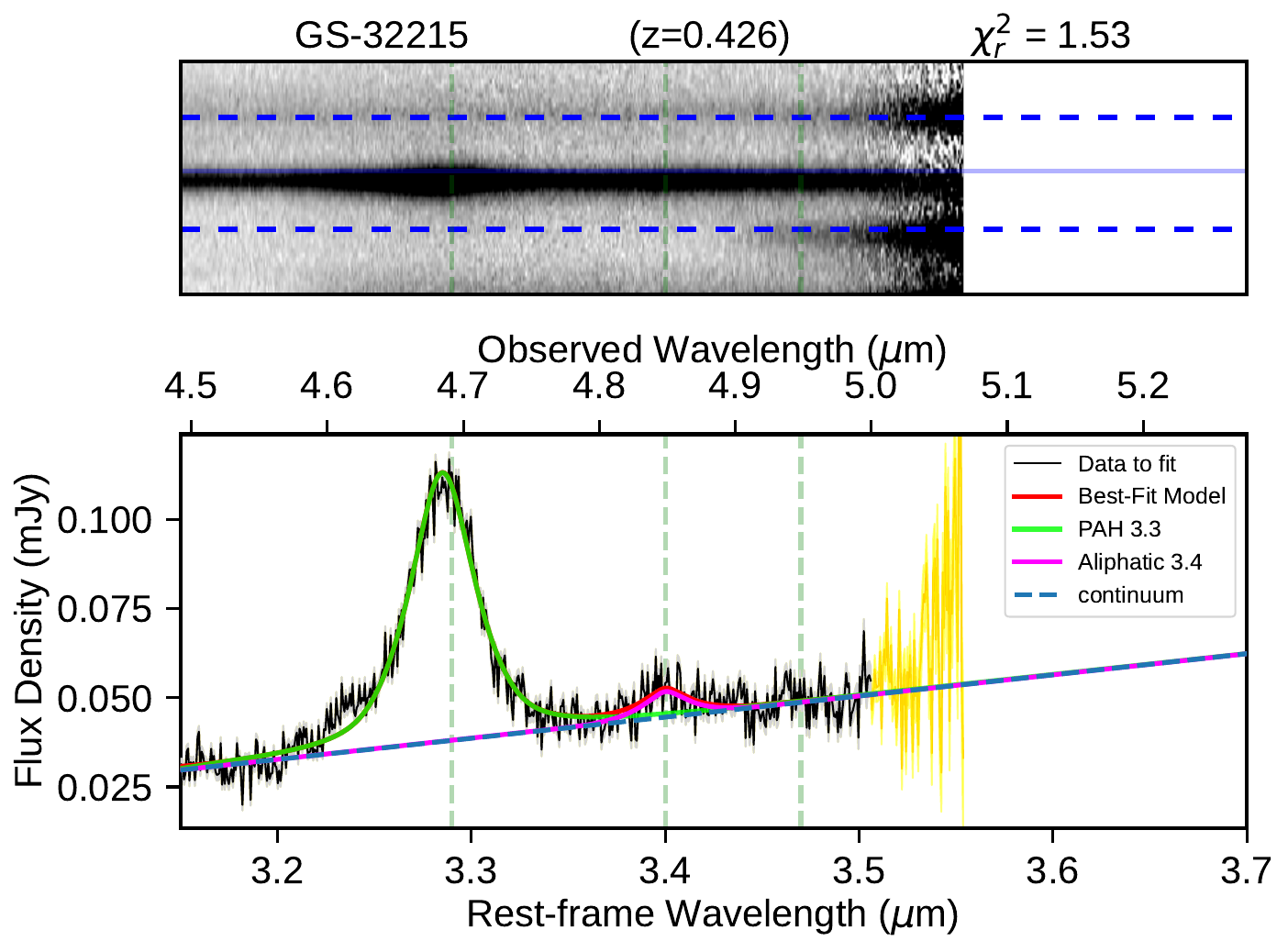}%
\includegraphics[width=0.495\hsize]{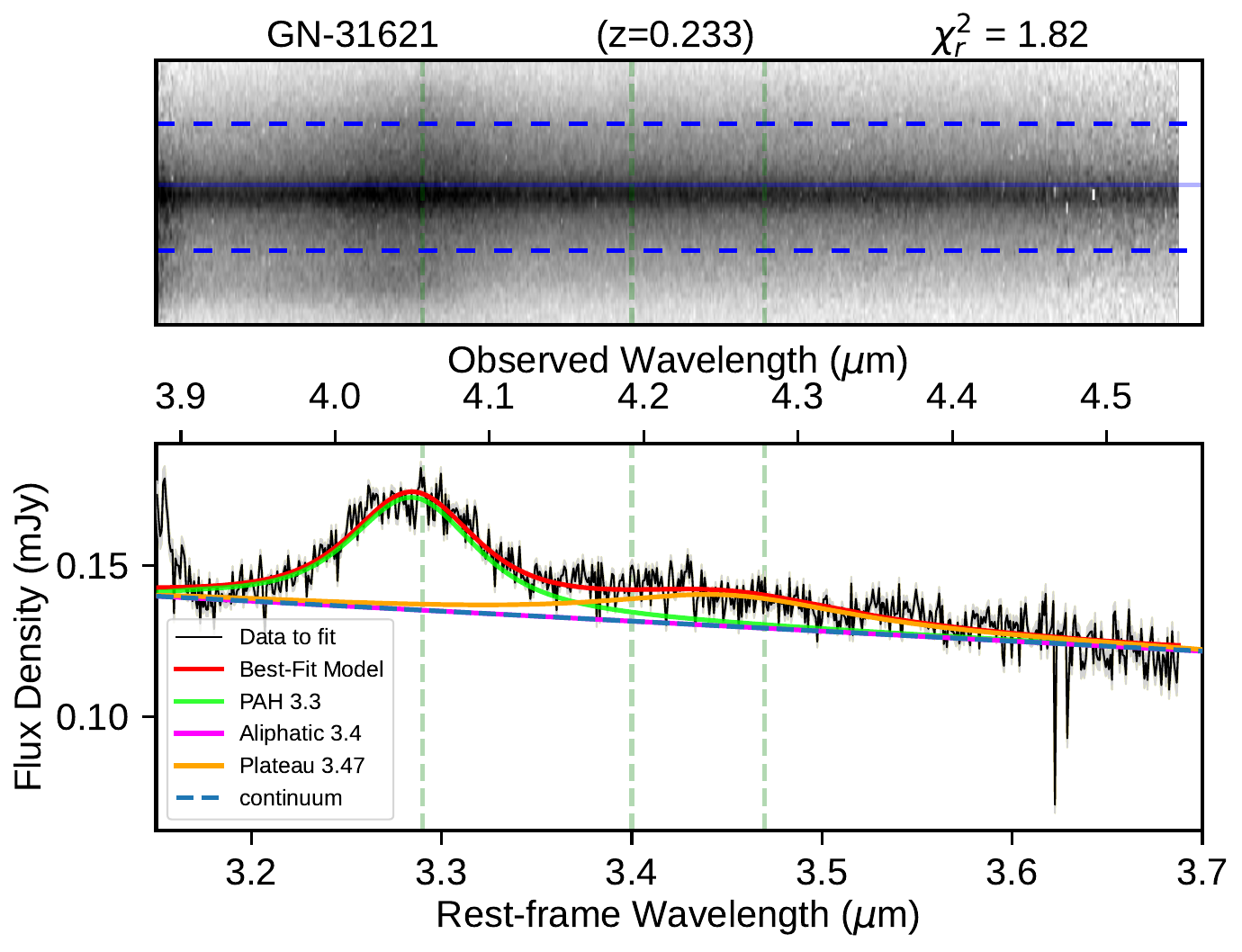}%
\\
\includegraphics[width=0.495\hsize]{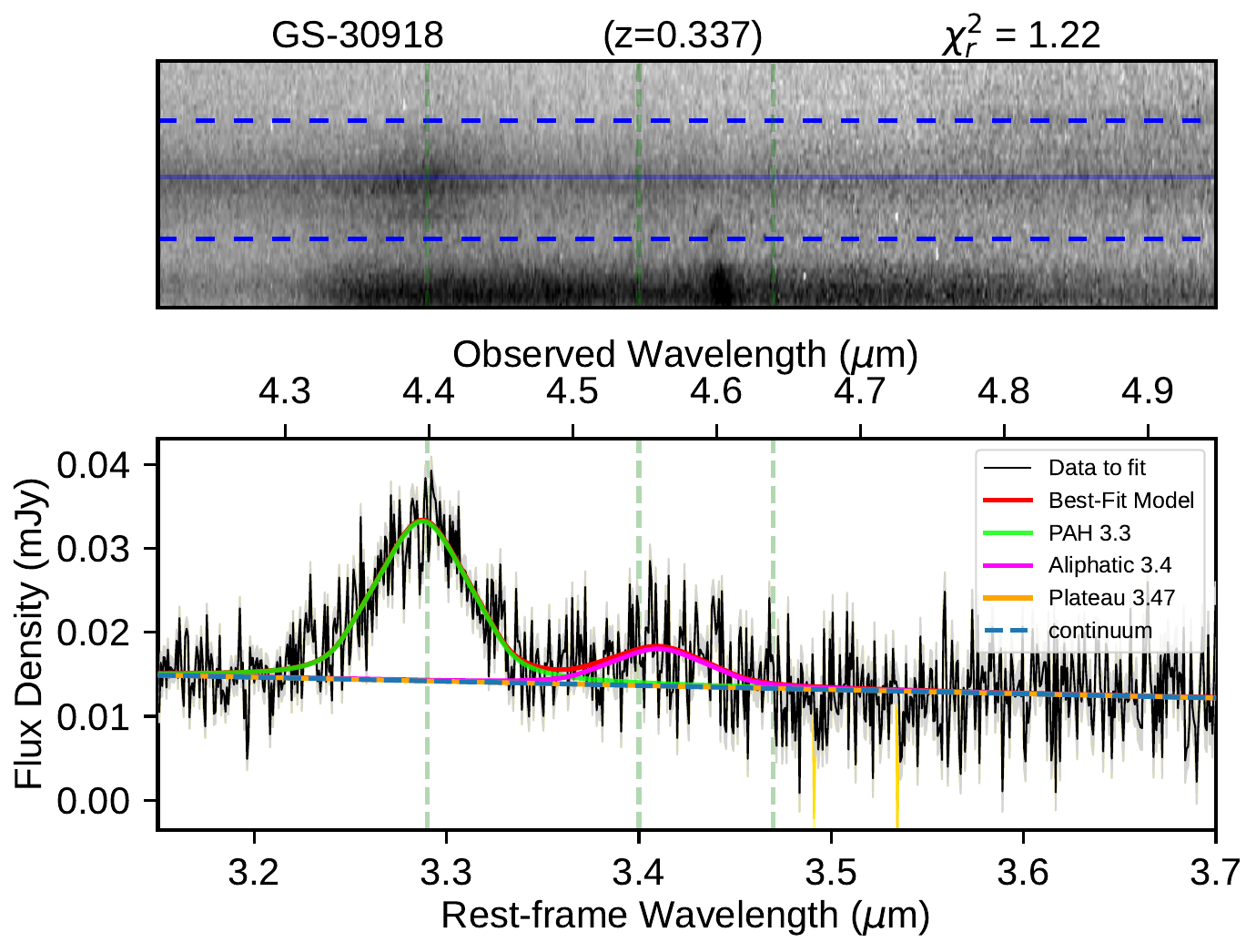}%
\includegraphics[width=0.495\hsize]{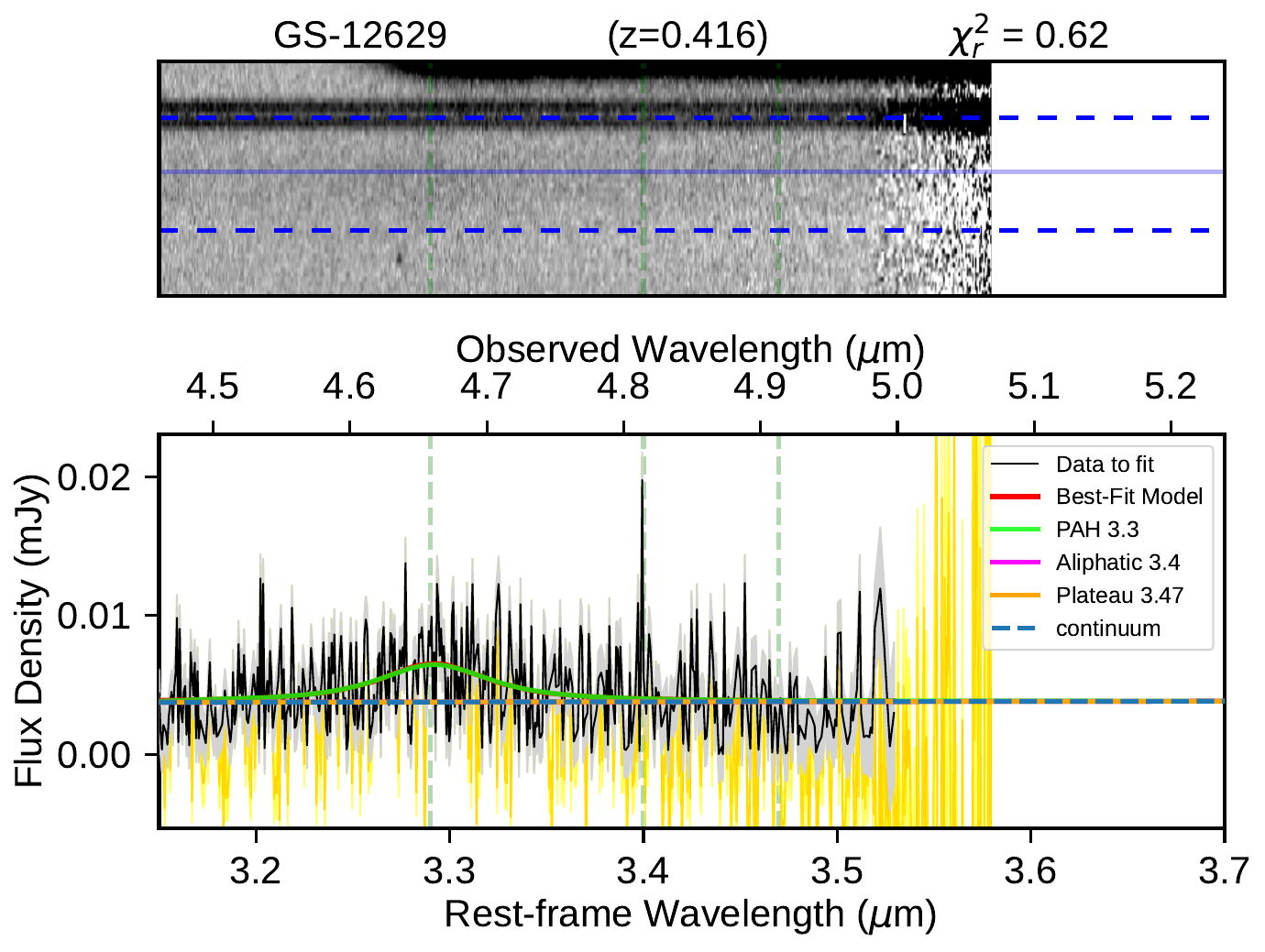}%
\\
\caption{\footnotesize  
         \label{fig:pah-spec}
         Example FRESCO 2-D spectral images (top) and extracted 1-D spectra (bottom) of
         example GOODS galaxies with 3.3$\mum$ PAH emission detections. For the
         2-D spectral images, we highlight the source center (solid blue line) and $\pm$0.75\arcsec distance along the dispersion direction (dashed blue line). In the spectral plot, the data used to constrain the model is shown in black lines with gray shades for the 1-$\sigma$ flux uncertainties. The orange line with yellow shades are the data and uncertainties dropped for the fittings. The best-fit model
         for the 1-D spectrum is denoted as a red solid line with the 3.3 $\mum$ component in green, 3.4 $\mum$ component in magenta, 3.47 $\mum$ plateau in orange and the featureless continuum in blue dashed line. The complete figure set (200 images) is available in the online journal. 
         }
\end{figure*}

In Figure~\ref{fig:pah-spec}, we show example FRESCO NIRCam/WFSS spectra and our
PAH model fittings for galaxies with different levels of 3.3$\mum$ PAH emission
detections. In Table~\ref{tab:pah-measure}, we present the measurements of the
PAH emission features from the spectral analysis. In total, we have made
3-$\sigma$ detections of the 3.3$\mum$ feature in 88 galaxies down to
$\simali$$2.5\times10^{-16}\erg\s^{-1}\cm^{-2}$ and the 3.4$\mum$ feature in 37
galaxies down to $\simali$$1.8\times10^{-16}\erg\s^{-1}\cm^{-2}$. Sources
without detections are mostly due to their weak star formation or diffuse
emission with the corresponding PAH emission falling below the detection limits.


\figsetstart
\figsetnum{2}
\figsettitle{Figure 2}

\figsetgrpstart
\figsetgrpnum{2.7}
\figsetgrptitle{FRESCO NIRCam/WFSS spectral fitting of GS-10260}
\figsetplot{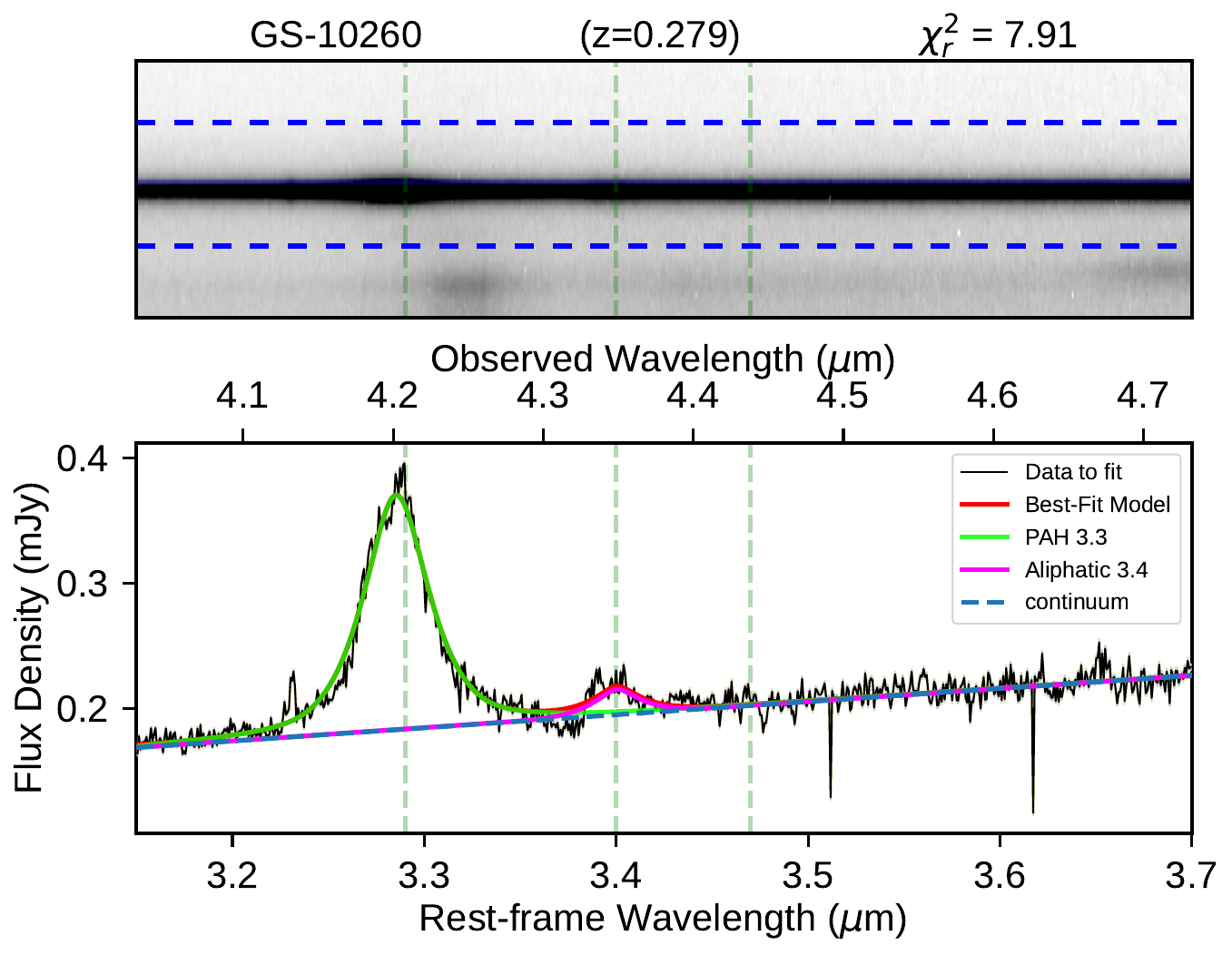}
\figsetgrpnote{FRESCO 2-D spectral image (top) and extracted 1-D spectrum (bottom) of the galaxy with 3.3$\mum$ PAH emission detections. For the
2-D spectral image, we highlight the source center (solid blue line) and $\pm$0.75\arcsec distance along the dispersion direction (dashed blue line). In the spectral plot, the data used to constrain the model is shown in black lines with gray shades for the 1-$\sigma$ flux uncertainties. The orange line with yellow shades are the data and uncertainties dropped for the fittings. The best-fit model
for the 1-D spectrum is denoted as a red solid line with the 3.3 $\mum$ component in green, 3.4 $\mum$ component in magenta, 3.47 $\mum$ plateau in orange and the featureless continuum in blue dashed line. }
\figsetgrpend

\figsetgrpstart
\figsetgrpnum{2.14}
\figsetgrptitle{FRESCO NIRCam/WFSS spectral fitting of GS-12629}
\figsetplot{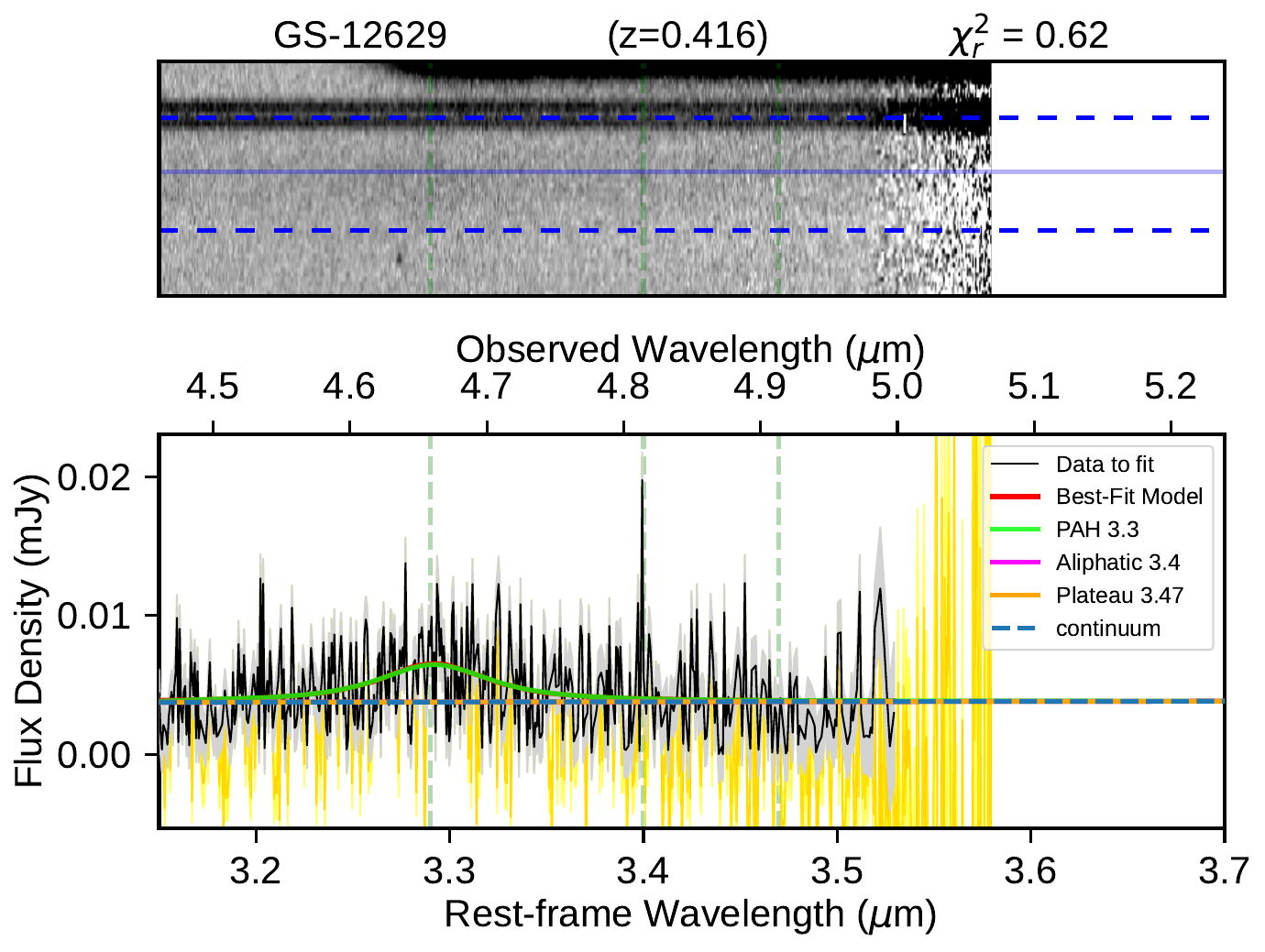}
\figsetgrpnote{FRESCO 2-D spectral image (top) and extracted 1-D spectrum (bottom) of the galaxy with 3.3$\mum$ PAH emission detections. For the
2-D spectral image, we highlight the source center (solid blue line) and $\pm$0.75\arcsec distance along the dispersion direction (dashed blue line). In the spectral plot, the data used to constrain the model is shown in black lines with gray shades for the 1-$\sigma$ flux uncertainties. The orange line with yellow shades are the data and uncertainties dropped for the fittings. The best-fit model
for the 1-D spectrum is denoted as a red solid line with the 3.3 $\mum$ component in green, 3.4 $\mum$ component in magenta, 3.47 $\mum$ plateau in orange and the featureless continuum in blue dashed line. }
\figsetgrpend

\figsetgrpstart
\figsetgrpnum{2.39}
\figsetgrptitle{FRESCO NIRCam/WFSS spectral fitting of GS-18467}
\figsetplot{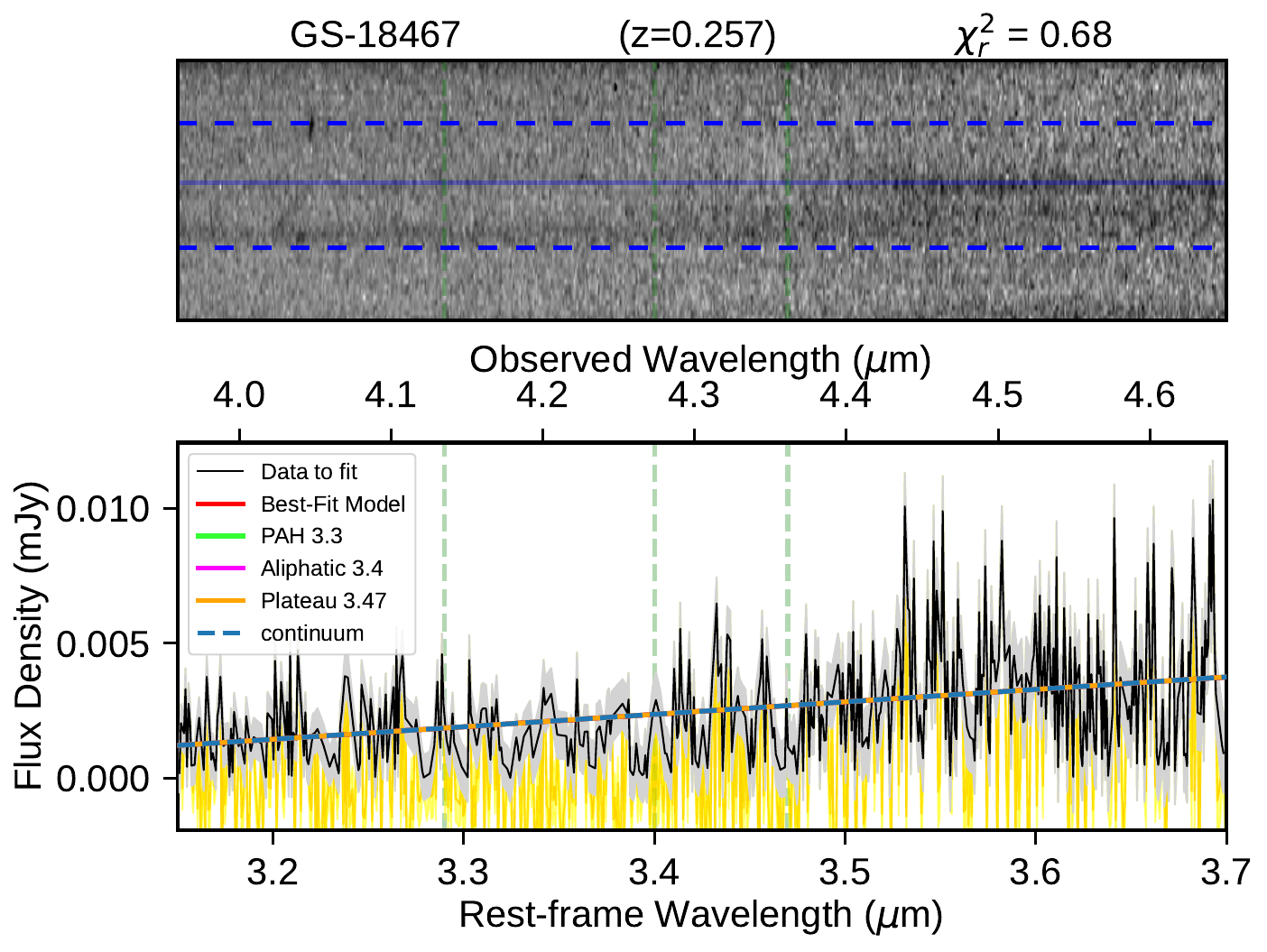}
\figsetgrpnote{FRESCO 2-D spectral image (top) and extracted 1-D spectrum (bottom) of the galaxy with 3.3$\mum$ PAH emission detections. For the
2-D spectral image, we highlight the source center (solid blue line) and $\pm$0.75\arcsec distance along the dispersion direction (dashed blue line). In the spectral plot, the data used to constrain the model is shown in black lines with gray shades for the 1-$\sigma$ flux uncertainties. The orange line with yellow shades are the data and uncertainties dropped for the fittings. The best-fit model
for the 1-D spectrum is denoted as a red solid line with the 3.3 $\mum$ component in green, 3.4 $\mum$ component in magenta, 3.47 $\mum$ plateau in orange and the featureless continuum in blue dashed line. }
\figsetgrpend

\figsetgrpstart
\figsetgrpnum{2.74}
\figsetgrptitle{FRESCO NIRCam/WFSS spectral fitting of GS-30918}
\figsetplot{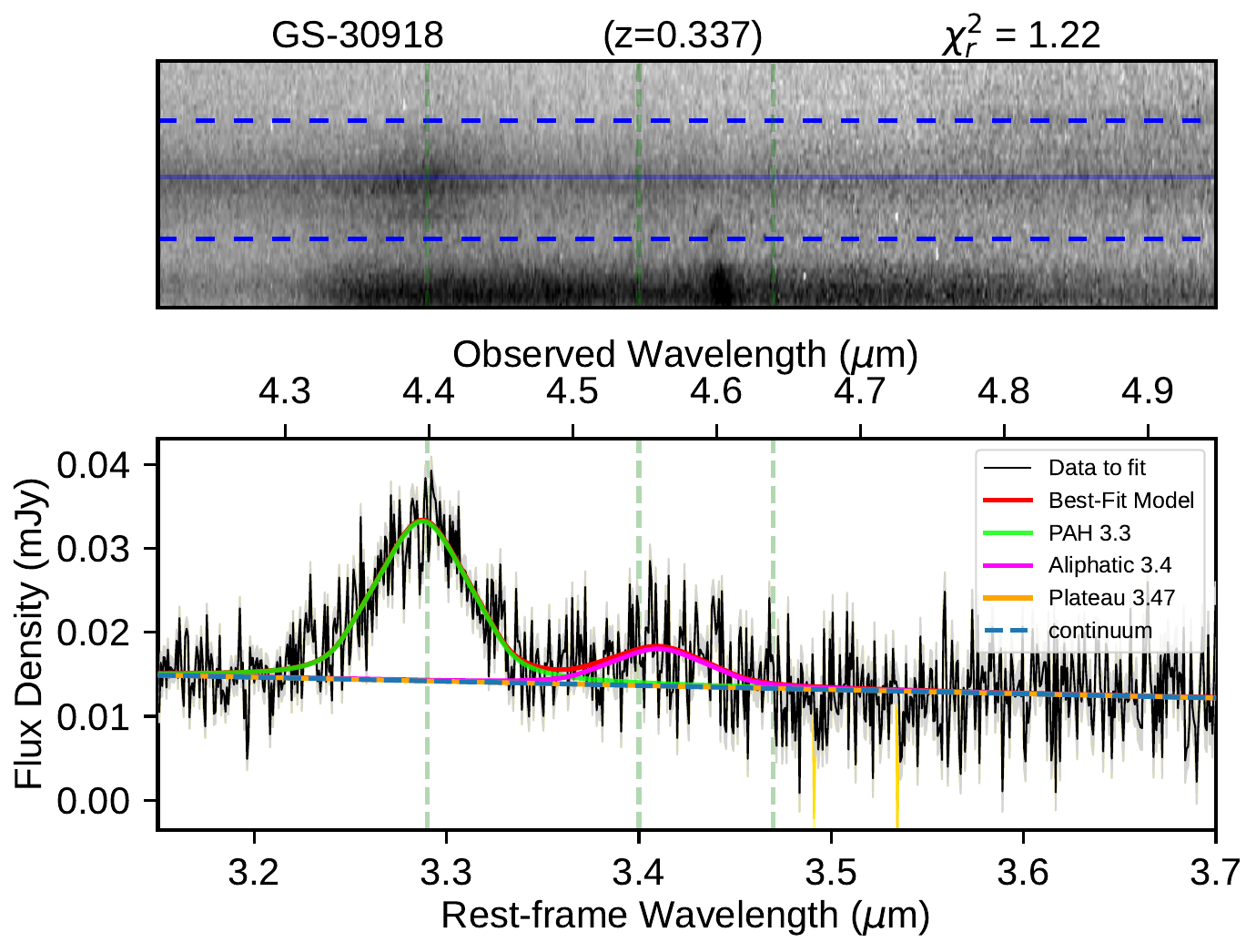}
\figsetgrpnote{FRESCO 2-D spectral image (top) and extracted 1-D spectrum (bottom) of the galaxy with 3.3$\mum$ PAH emission detections. For the
2-D spectral image, we highlight the source center (solid blue line) and $\pm$0.75\arcsec distance along the dispersion direction (dashed blue line). In the spectral plot, the data used to constrain the model is shown in black lines with gray shades for the 1-$\sigma$ flux uncertainties. The orange line with yellow shades are the data and uncertainties dropped for the fittings. The best-fit model
for the 1-D spectrum is denoted as a red solid line with the 3.3 $\mum$ component in green, 3.4 $\mum$ component in magenta, 3.47 $\mum$ plateau in orange and the featureless continuum in blue dashed line. }
\figsetgrpend

\figsetgrpstart
\figsetgrpnum{2.81}
\figsetgrptitle{FRESCO NIRCam/WFSS spectral fitting of GS-32215}
\figsetplot{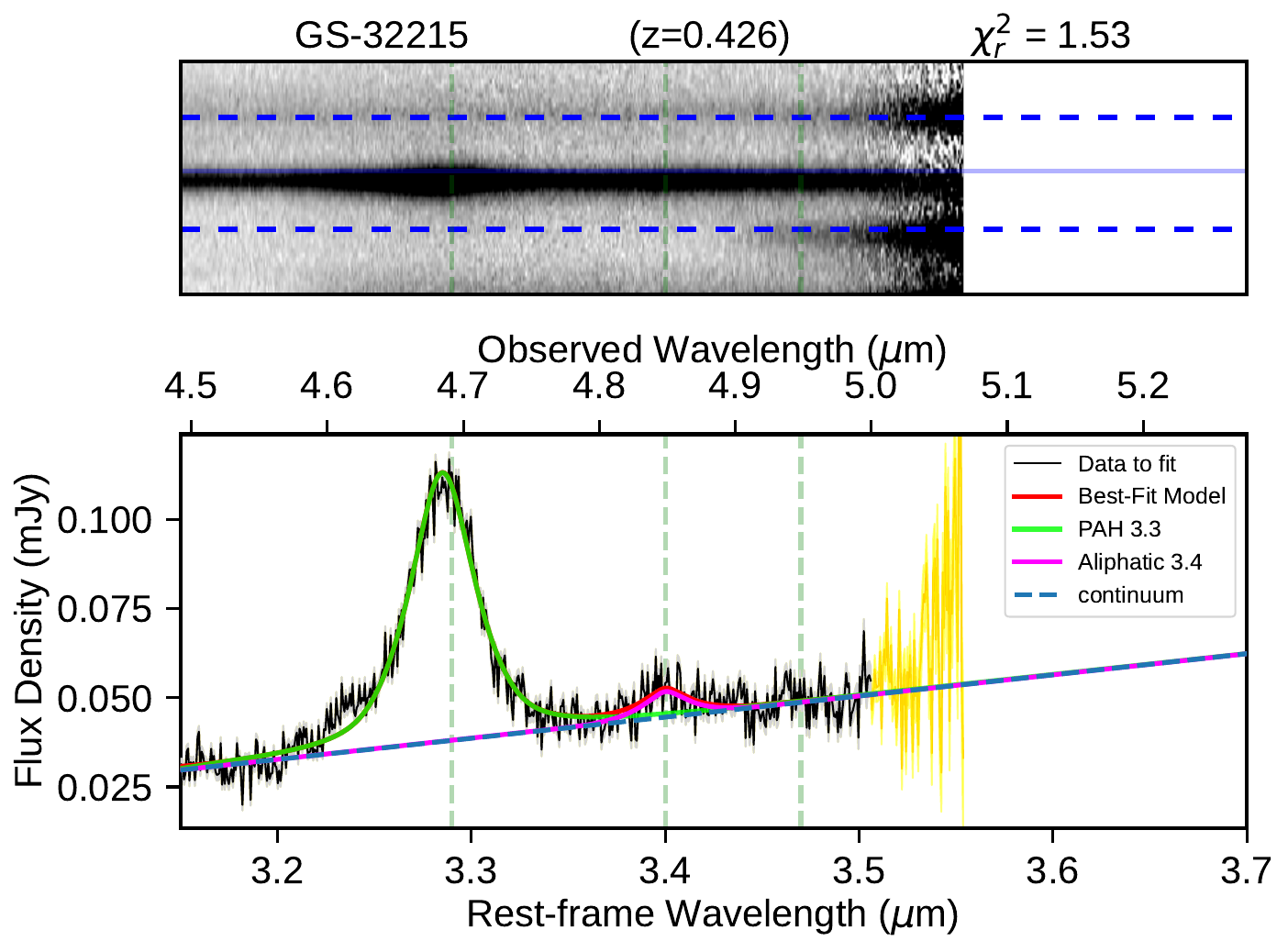}
\figsetgrpnote{FRESCO 2-D spectral image (top) and extracted 1-D spectrum (bottom) of the galaxy with 3.3$\mum$ PAH emission detections. For the
2-D spectral image, we highlight the source center (solid blue line) and $\pm$0.75\arcsec distance along the dispersion direction (dashed blue line). In the spectral plot, the data used to constrain the model is shown in black lines with gray shades for the 1-$\sigma$ flux uncertainties. The orange line with yellow shades are the data and uncertainties dropped for the fittings. The best-fit model
for the 1-D spectrum is denoted as a red solid line with the 3.3 $\mum$ component in green, 3.4 $\mum$ component in magenta, 3.47 $\mum$ plateau in orange and the featureless continuum in blue dashed line. }
\figsetgrpend

\figsetgrpstart
\figsetgrpnum{2.195}
\figsetgrptitle{FRESCO NIRCam/WFSS spectral fitting of GN-30345}
\figsetplot{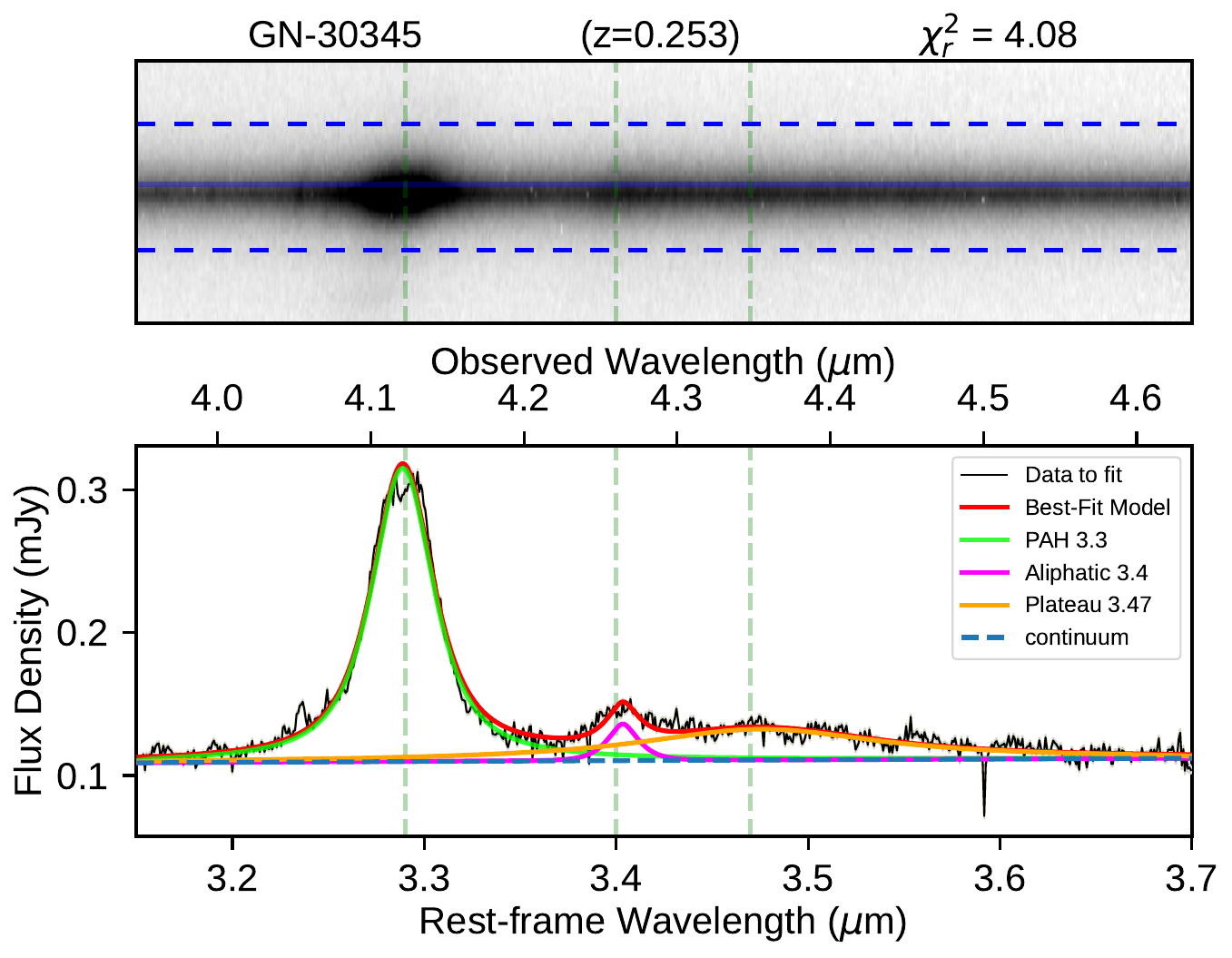}
\figsetgrpnote{FRESCO 2-D spectral image (top) and extracted 1-D spectrum (bottom) of the galaxy with 3.3$\mum$ PAH emission detections. For the
2-D spectral image, we highlight the source center (solid blue line) and $\pm$0.75\arcsec distance along the dispersion direction (dashed blue line). In the spectral plot, the data used to constrain the model is shown in black lines with gray shades for the 1-$\sigma$ flux uncertainties. The orange line with yellow shades are the data and uncertainties dropped for the fittings. The best-fit model
for the 1-D spectrum is denoted as a red solid line with the 3.3 $\mum$ component in green, 3.4 $\mum$ component in magenta, 3.47 $\mum$ plateau in orange and the featureless continuum in blue dashed line. }
\figsetgrpend

\figsetgrpstart
\figsetgrpnum{2.196}
\figsetgrptitle{FRESCO NIRCam/WFSS spectral fitting of GN-30796}
\figsetplot{GN_30796_pahfit.pdf}
\figsetgrpnote{FRESCO 2-D spectral image (top) and extracted 1-D spectrum (bottom) of the galaxy with 3.3$\mum$ PAH emission detections. For the
2-D spectral image, we highlight the source center (solid blue line) and $\pm$0.75\arcsec distance along the dispersion direction (dashed blue line). In the spectral plot, the data used to constrain the model is shown in black lines with gray shades for the 1-$\sigma$ flux uncertainties. The orange line with yellow shades are the data and uncertainties dropped for the fittings. The best-fit model
for the 1-D spectrum is denoted as a red solid line with the 3.3 $\mum$ component in green, 3.4 $\mum$ component in magenta, 3.47 $\mum$ plateau in orange and the featureless continuum in blue dashed line. }
\figsetgrpend

\figsetgrpstart
\figsetgrpnum{2.197}
\figsetgrptitle{FRESCO NIRCam/WFSS spectral fitting of GN-31621}
\figsetplot{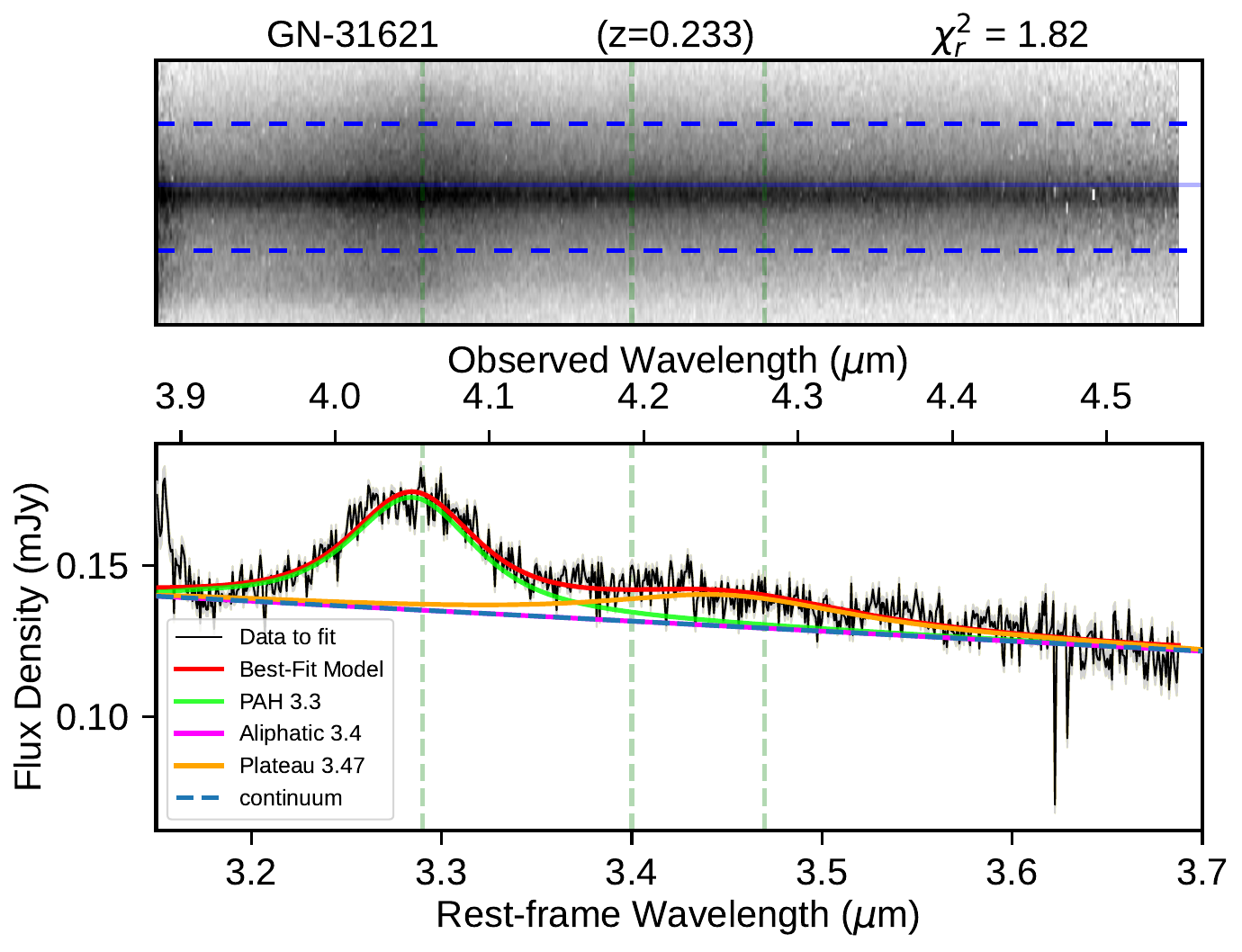}
\figsetgrpnote{FRESCO 2-D spectral image (top) and extracted 1-D spectrum (bottom) of the galaxy with 3.3$\mum$ PAH emission detections. For the
2-D spectral image, we highlight the source center (solid blue line) and $\pm$0.75\arcsec distance along the dispersion direction (dashed blue line). In the spectral plot, the data used to constrain the model is shown in black lines with gray shades for the 1-$\sigma$ flux uncertainties. The orange line with yellow shades are the data and uncertainties dropped for the fittings. The best-fit model
for the 1-D spectrum is denoted as a red solid line with the 3.3 $\mum$ component in green, 3.4 $\mum$ component in magenta, 3.47 $\mum$ plateau in orange and the featureless continuum in blue dashed line. }
\figsetgrpend

\figsetend

\begin{deluxetable*}{cccccccccc}
\tablecaption{{Measurements of 3.3 $\mum$ aromatic and 3.4$\mum$ aliphatic features for galaxies in the FRESCO footprint}}
\tablehead{
\colhead{ID} &
\colhead{$z$} &
\colhead{$D$} &
\colhead{$\lambda_{3.3}$} &
\colhead{$P_{3.3}$} &
\colhead{${\rm tag}_{3.3}$} &
\colhead{$\lambda_{3.4}$} &
\colhead{$P_{3.4}$} &
\colhead{${\rm tag}_{3.4}$}  &
\colhead{$\alifrac$} \\
\colhead{} &
\colhead{} &
\colhead{(cm)} &
\colhead{($\mum$)} &
\colhead{(erg$\s^{-1}\cm^{-2}$)} &
\colhead{} &
\colhead{($\mum$)} &
\colhead{(erg$\s^{-1}\cm^{-2}$)} &
\colhead{}  &
\colhead{(\%)} 
}
\startdata
GS-08629 &	0.279 &  4.46E+27      &	3.284$\pm$0.000 & 	2.10E-15$\pm$6.08E-18 &	1	& 3.400$\pm$0.001  &	4.41E-16$\pm$4.22E-17&	1  & 3.18  \\
GS-09402 &	0.359 &  5.99E+27      &	3.283$\pm$0.001 &	1.85E-16$\pm$9.57E-18 &	1	& 3.416$\pm$0.014  & 	 $<$5.95E-19	     &  1  &  0.0   \\
GS-09755 &	0.413 &  7.06E+27      &	3.317$\pm$0.011	&   1.22E-15$\pm$8.54E-17 &	1   &  N/A.   	       &      N.A.      		 &  0  & N.A.   \\
GS-09883 &	0.347 &  6.08E+27      &	3.290$\pm$0.000	&   5.60E-16$\pm$3.19E-17 &	1	& 3.406$\pm$0.010  &  4.75E-17$\pm$4.04E-17	 &  1  &  1.31 \\
$\cdots$ & $\cdots$ & $\cdots$   & $\cdots$ &  $\cdots$  &  $\cdots$  &  $\cdots$  &   $\cdots$   &  $\cdots$   &  $\cdots$ \\
\enddata
\tablecomments{Here $D$ represent the luminosity distance of the galaxy, ${\rm tag}_{3.3}$ 
and ${\rm tag}_{3.4}$ indicate whether 
the reported measurements can be trusted 
(${\rm tag}=1$) or not (${\rm tag}=0$). 
Only a portion of this table is shown here
to demonstrate its form and content. 
A machine-readable version of 
the full table is available online.}
\label{tab:pah-measure}
\end{deluxetable*}

\section{Results and Discussion}
With the measurements of the 3.3 and 3.4$\mum$ features in hand, we now
explore their behaviors across the sample and assess their connection to galaxy
properties.

\subsection{The 3.3$\mum$ Aromatic Band}
\subsubsection{$L_{3.3}$ vs. $L_{\rm IR}$}
A key starting point is to assess the relationship between PAH emission and the
total IR luminosity, a well-established tracer of the overall dust heating in
galaxies. This analysis provides the foundation for understanding how PAH
features evolve across different environments.

\begin{figure}
\begin{center}
\includegraphics[width=1.0\hsize]{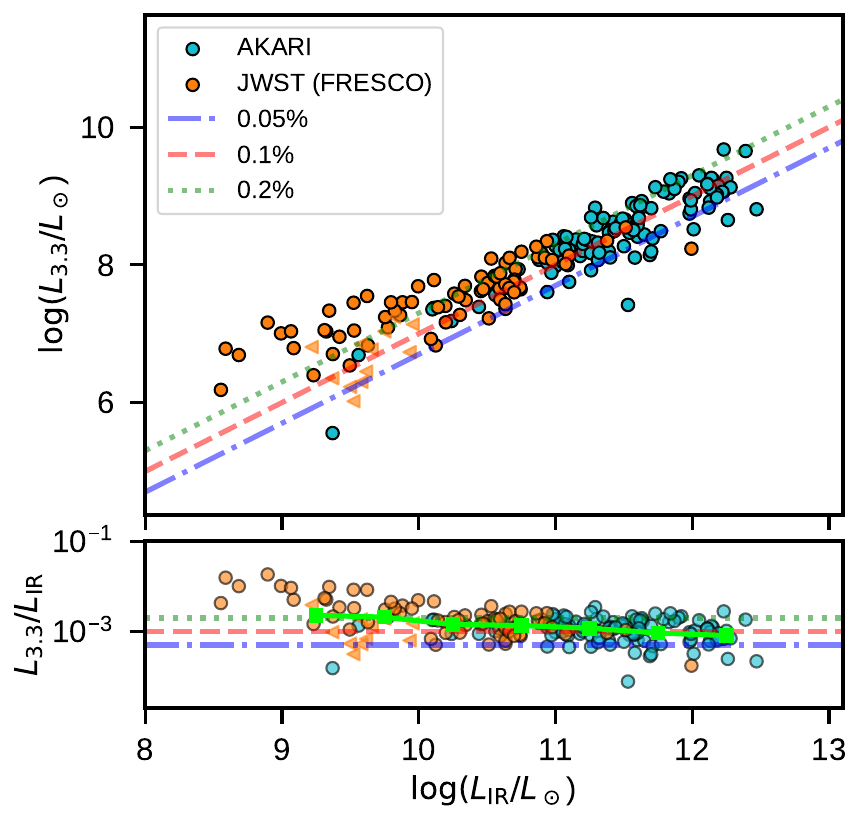}
\end{center}
\caption{
Top panel: Correlation between the galaxy
3.3$\mum$ PAH luminosity ($L_{3.3}$)
and IR luminosity ($\LIR$),
from our sample (orange dots) and
those obtained with AKARI (blue dots).
Bottom panel: Variation of $L_{3.3}/\LIR$ with $\LIR$.
In both plots, $L_{3.3}/\LIR$\,=\,0.0005, 0.001,
and 0.002 are shown as dot-dashed, dashed and
dotted lines, respectively. In bottom panel,
the green-square-connected solid line plots
the averaged $L_{\rm 3.3}/\LIR$ value
in different bins of $\LIR$. See text for details.
         }
\label{fig:lpah_lir}
\end{figure}

In Figure~\ref{fig:lpah_lir}, we compare the IR luminosity of the galaxies in
our sample with their 3.3$\mum$ PAH luminosity ($L_{3.3} =
P_{3.3}\times4\pi\times d^2$, where $d$ is the galaxy luminosity distance). To
extend the dynamical range, we include data from the brighter IR galaxies
observed with AKARI \citep{Imanishi2010, Ichikawa2014}. A strong correlation is
evident with a Pearson correlation coefficient of $r\approx0.71$ and $p$-value
on the order of $10^{-32}$. The Spearman test yields $\rho\approx0.94$ with a
$p$-value of $\simali$$10^{-92}$. These results indicate a robust and
statistically significant link betwen $L_{3.3}$ and $L_{\rm IR}$, suggesting
both types of emission originate from heating by UV photons in star-forming
regions. 

We also found tentative evidence that the ratio $L_{3.3}/\LIR$ may change from $\simali$0.01\% at
$\LIR$\,$\simali$$10^{11}$--$10^{12.5}\Lsun$ to $\simali$0.02\% at
$\LIR$\,$\simali$$10^{9}$--$10^{10}\Lsun$. { A similar trend has been previously reported by \cite{Yamada2013} based on AKARI observations of 101 IR-bright galaxies at $z\lesssim0.1$ with a much smaller sample of just 13 objects at $\LIR\lesssim10^{11}~\Lsun$ and 2 objects at $\LIR\lesssim10^{10}$.} Such a trend mirrors the observed
behavior of the other PAH bands, such as the 11.3$\mum$ feature \citep{Lyu2017}.
Nevertheless, the change of the relative
intensity of the 3.3$\mum$ feature is only moderate within the three orders of
magnitude variation of $\LIR$, supporting $L_{3.3}$ as a valid tracer of star
formation as discussed below.

\subsubsection{$L_{3.3}$ vs. SFR}
\begin{figure}
\begin{center}
\includegraphics[width=1.0\hsize]{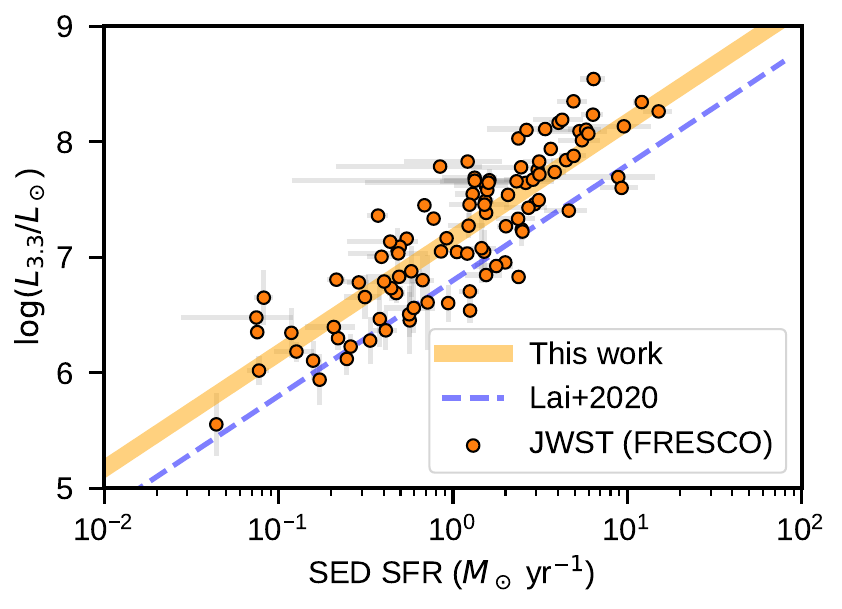}
\end{center}
\caption{Top panel: Relation between the galaxy 3.3$\mum$
PAH luminosity $L_{3.3}$ and the SED-derived SFR.
The orange solid thick line is the fitted correlation in log-log space
from this work, and the blue dashed thin line is the
$L_{3.3}$--SFR relation presented in \citet{Lai2020}.
         }
         \label{fig:lpah_sfr}
\end{figure}

Building on the strong correlation
between $L_{3.3}$ and $\LIR$,
we now evaluate how the 3.3$\mum$ PAH feature
serves as a tracer of star formation activity in galaxies.
Adopting the $L_{\rm IR}$-calibrated SFR 
\citep[e.g.,][]{kennicutt2012} and the average ratio of
$\log\,\left(L_{3.3}/\LIR\right)\,=\,-2.77\pm0.30$,
we can compute the (IR-based) SFR
from the 3.3$\mum$ PAH luminosity as 
\begin{equation}
\log\,\left(\frac{\rm SFR}{\Msun\yr^{-1}}\right)
= \log\,\left(\frac{L_{3.3}}{\Lsun}\right) - \left(7.03\pm0.30\right)~~.
\label{equ:lir-pah}
\end{equation}
Meanwhile, we can also use the SED-derived SFR to examine its relation with
$L_{3.3}$. As shown in Figure~\ref{fig:lpah_sfr}, these two independent
quantities exhibit a strong correlation. We conduct linear regression analysis
in the log-log space with a fixed slope of 1.0 and find the following
correlation:
\begin{equation}
  \log\,\left(\frac{L_{3.3}}{\Lsun}\right)
  \,=\,\log\,\left(\frac{\rm SFR}{\Msun\yr^{-1}}\right)
\,+\,\left(7.17\pm0.32\right)~~.
\end{equation}
The scatter is $\simali$0.37. Considering the typical uncertainties of 0.2--0.3
dex of SFR from {\sf Prospector}, the $L_{3.3}$--SFR correlation has a
$\simali$0.2--0.3 dex intrinsic scatter, indicating a tight correlation between
$L_{3.3}$ and the SED-derived SFR. The same relation can be rewritten as 
\begin{equation}
\log\,\left(\frac{\rm SFR}{\Msun\yr^{-1}}\right)
= \log\,\left(\frac{L_{3.3}}{\Lsun}\right) - \left(7.17\pm0.32\right)~~.
\label{equ:sed-pah}
\end{equation}
This correlation should be more robust since that the SFR based on SED analysis considers both obscured and unobscured stellar population and the validity of this approach is well-established regardless of the galaxy IR luminosity or stellar mass.
In comparison,  the $\LIR$-based SFR calibration in Equation~\ref{equ:lir-pah} is slightly off by $\simali$0.14 dex but still within the
uncertainty. 
%

{ Based on AKARI and {\it Spitzer} data,} \citet{Lai2020} calibrated the 3.3$\mum$ PAH luminosity
as a SFR indicator by comparing it to the mid-IR [Ne\,II]
and [Ne\,III] lines. Compared to that work, for the same
3.3$\mum$ PAH luminosity, the SED-calibrated SFR
and the $L_{\rm IR}$-based SFR would be $\simali$0.37 dex
and 0.23 dex lower, respectively. 
These offsets might be explained as a result of 
systematic differences in SFR calibration,
adopted methodology as well as various measurement
uncertainties. For example, the 3.3$\mum$
PAH emission (relative to the IR luminosity) is boosted by
a factor of $\simali$2 (0.3 dex) in the low IR luminosity
range studied here compared to the high luminosity
range in \citet{Lai2020}.
Nevertheless, considering the large calibration uncertainties
in these correlations ($\simali$0.3--0.5 dex),
such offsets are not very significant,
indicating a general agreement.
In other words, the 3.3$\mum$ PAH luminosity
does effectively trace star formation rates
across a wide range from $\simali$0.1~$M_\odot\yr^{-1}$
to $\simali$300~$M_\odot\yr^{-1}$.

\subsubsection{$L_{3.3}$ vs. Metallicity}\label{sec:L3.3_Metal}
Given the role of metallicity in regulating PAH formation
and survival, we also explore the dependence of
$L_{3.3}/\LIR$ on the gas-phase metallicity,
probing potential links between the chemical
enrichment and the PAH emission efficiency.

\begin{figure}
\begin{center}
\includegraphics[width=1.0\hsize]{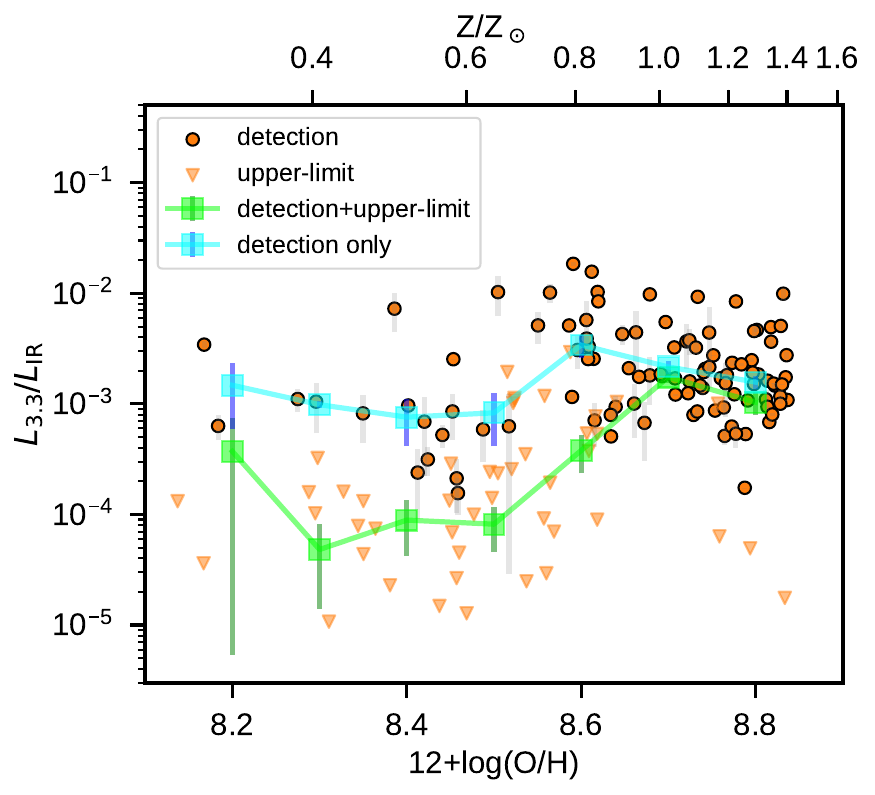}
\end{center}
\caption{The galaxy 3.3$\mum$ PAH emission
relative to the total IR luminosity ($L_{3.3}/\LIR$)
as a function of galaxy metallicity $12+\log({\rm O/H})$
inferred from the FMR. We computed the average values
of $L_{3.3}/\LIR$ in different bins of $12+\log({\rm O/H})$
for sources with 3.3$\mum$ PAH detections only
(blue squares) and sources with both detections
and upper limits (green squares). To compute mean values with measurements that include upper limits, we used the Kaplan-Meier analysis. 
         }
         \label{fig:lpah_metal}
\end{figure}

In Figure~\ref{fig:lpah_metal}, we show
the relative strength of the 3.3$\mum$ PAH emission
to the galaxy IR luminosity as a function of galaxy metallicity
$12+\log\left({\rm O/H}\right)$ estimated based on FMR.
The correlation analysis between $L_{3.3}/\LIR$
and metallicity $12+\log({\rm O/H})$ shows a weak
but statistically significant monotonic relationship.
The Spearman correlation ($\rho\approx0.21$, $p\approx0.032$)
and Kendall's tau ($\tau\approx0.14$, $p\approx0.043$)
indicate that as the metallicity increases,
$L_{3.3}/\LIR$ tends to increase, though the trend is weak.
After binning the data points by metallicity,
we can clearly see a drop at lower metallicities,
regardless of whether the non-detections
are considered or not. The change happens at
$12+\log({\rm O/H})$\,$\simali$8.4--8.5,
corresponding to $\simali$60\%--80\% solar metallicity.
This trend is roughly consistent with previous studies
of other PAH bands in the mid-IR obtained
with {\it Spitzer} such as  \cite{Draine2007,  Marble2010}
and \cite{Whitcomb2024}, including galaxies
at redshifts up to $\simali$2 \citep{Shivaei2024}.

The exact reason for the deficiency of PAHs
in low-metallicity galaxies is not clear
(see \citealt{Li2020}). It is generally interpreted as more
rapid destruction of PAHs by the more intense
and harder UV radiation
(as indicated by the fine-structure line ratio of 
[Ne{\sc iii}]/[Ne{\sc ii}])
in an ISM with reduced shielding by dust.
Low-metallicity environments lack sufficient
dust grains to shield PAHs from photodissociation
by UV radiation. 
It could also be due to more effective destruction of PAHs 
by thermal sputtering in shock-heated gas that cools more
slowly because of the reduced metallicity (see \citealt{Li2020}).

The trend of $L_{3.3}/\LIR$ increasing
with metallicity can also be explained 
by the delayed enrichment of PAHs in the ISM,
as highlighted by \citet{Galliano2008}.
In low-metallicity galaxies, the primary contributors
to PAH production---carbon-rich asymptotic giant
branch (AGB) stars---take hundreds of millions of
years to evolve and inject PAHs into the ISM.
As a result, young, metal-poor galaxies experience
a deficit in PAH emission,
leading to lower $L_{3.3}/\LIR$ ratios.
In contrast, higher-metallicity galaxies,
which have undergone more extensive star formation
and stellar evolution, benefit from the cumulative
contribution of AGB stars, enriching the ISM
with PAHs over time. This enrichment boosts
the PAH emission relative to the total IR luminosity.
The increasing $L_{3.3}/\LIR$ ratio at higher metallicities
reflects the more developed carbon chemistry
in these galaxies, where PAH formation
and survival are favored. This evolutionary process,
coupled with the reduced destruction of PAHs
in less harsh radiation environments of mature galaxies,
naturally explains the observed metallicity dependence
of $L_{3.3}/\LIR$.
%

\subsection{The 3.4$\mum$ Aliphatic Band}
As mentioned earlier, the 3.4$\mum$ emission band
provides insight into the aliphatic component of
PAH molecules. We now turn to the analysis of
this weaker yet crucial feature to further explore
the composition and processing of PAHs in galaxies.

For galaxies with both the 3.3$\mum$ aromatic
and 3.4$\mum$ aliphatic features detected
at 3-$\sigma$ significance, a large variation
of the aliphatic-to-aromatic band ratio 
($P_{3.4}/P_{3.3}$) is observed from 
$\simali$0.05 to $\simali$0.58,
with a mean value of $\simali$0.22
(and a standard deviation of $\simali$0.14)
and a median value of $\simali$0.19.
As discussed in \citet{yang2023},
the aliphatic fractions ($\alifrac$) of PAHs,
defined as the fractions of C atoms
in aliphatic units, can be determined
from the 3.4$\mum$-to-3.3$\mum$
emission intensity ratios ($P_{3.4}/P_{3.3}$):
\begin{equation}\label{eq:alifrac1}
\alifrac = \left(1+\NCaro/\NCali\right)^{-1} ~~,
\end{equation}
\begin{equation}\label{eq:alifrac2}
\frac{\NCali}{\NCaro}
\approx \frac{1}{6.40}
\left(\frac{P_{3.4}}{P_{3.3}}\right) ~~,
\end{equation}
where $\NCaro$ and $\NCali$ are respectively
the number of aromatic and aliphatic C atoms
in a PAH molecule.
In the FRESCO galaxies considered here,
the observed $P_{3.4}/P_{3.3}$ ratios
translate into aliphatic fractions of
$\alifrac$ from $\simali$0.78\% to $\simali$8.3\%,
with a mean value of $\simali$3.3\%
(and a standard deviation of $\simali$1.9\%)
and a median value of $\simali$2.9\%.
As the PAH aliphatic fraction $\alifrac$
is linearly related to $P_{3.4}/P_{3.3}$,
provided $\NCali\ll\NCaro$ 
(see eqs.\,\ref{eq:alifrac1},\,\ref{eq:alifrac2}),
in the following we will discuss the PAH
aliphacity simply in terms of 
$P_{3.4}/P_{3.3}$  or $L_{3.4}/L_{3.3}$,
where $L_{3.3} = P_{3.3}\times4\pi\times D^2$,
$L_{3.4} = P_{3.4}\times4\pi\times D^2$,
and $D$ is the luminosity distance to the galaxy.

\begin{figure*}
    \centering
    \includegraphics[width=0.32\hsize]{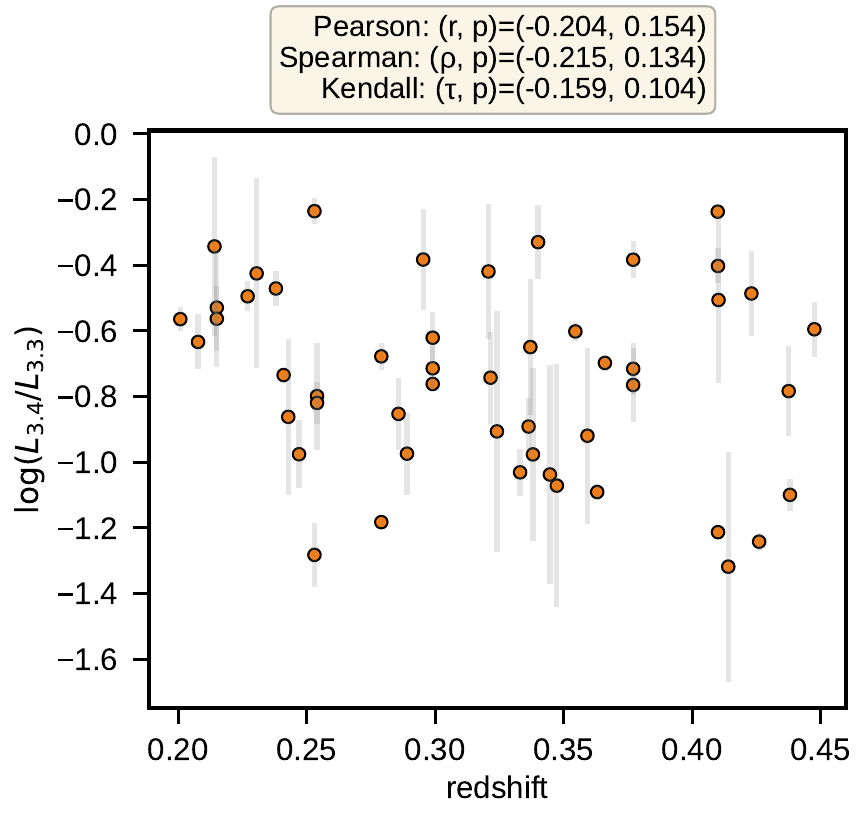}
    \includegraphics[width=0.32\hsize]{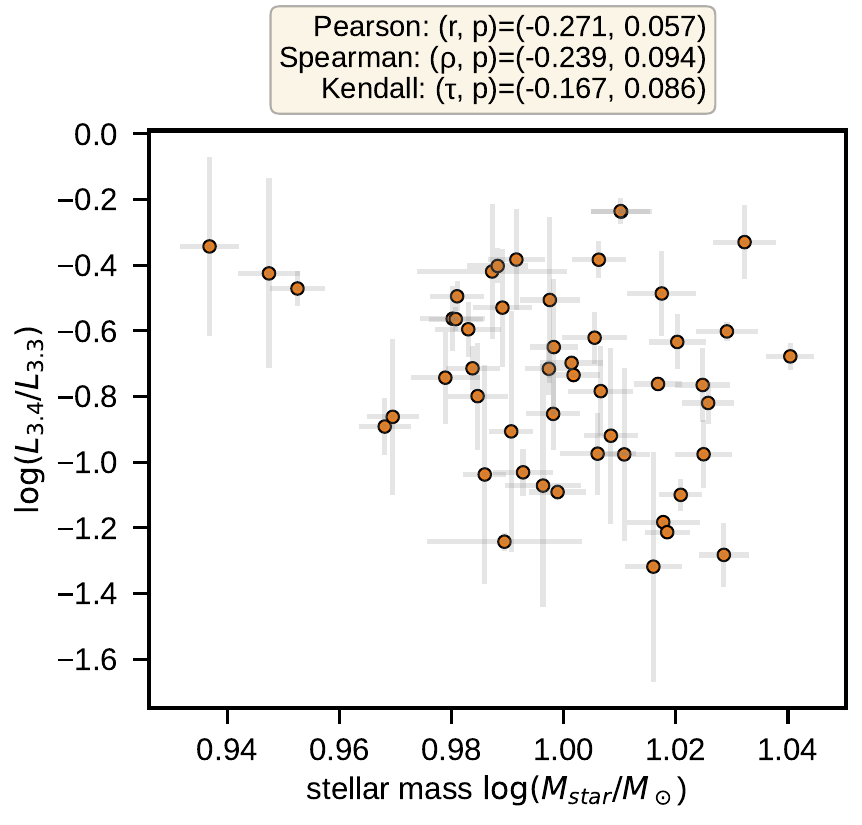}
    \includegraphics[width=0.32\hsize]{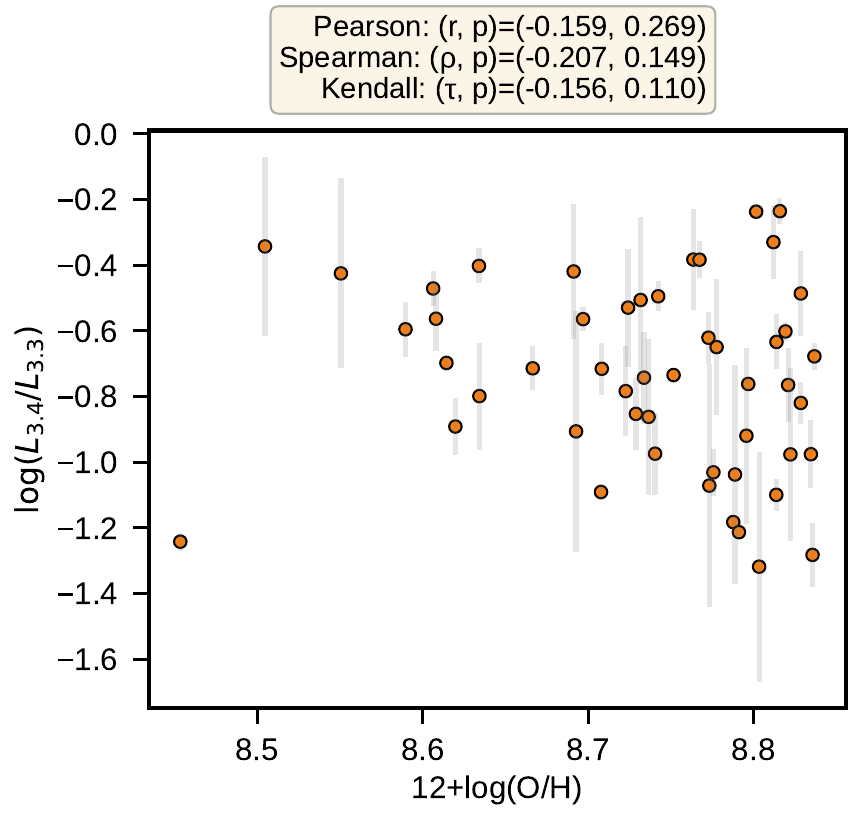}
    \includegraphics[width=0.32\hsize]{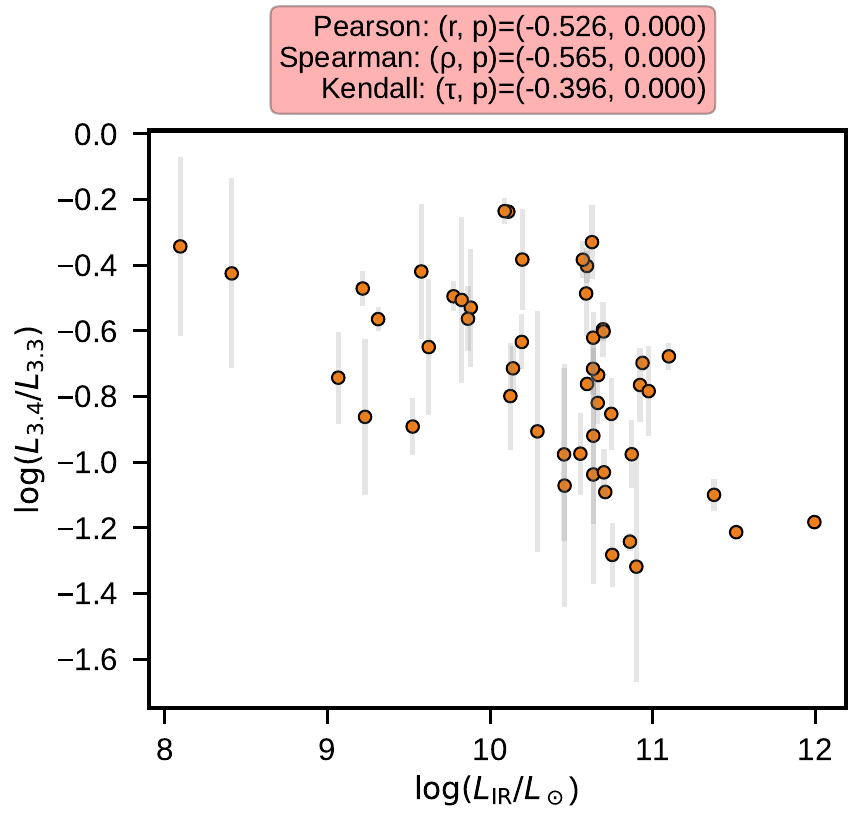}
    \includegraphics[width=0.32\hsize]{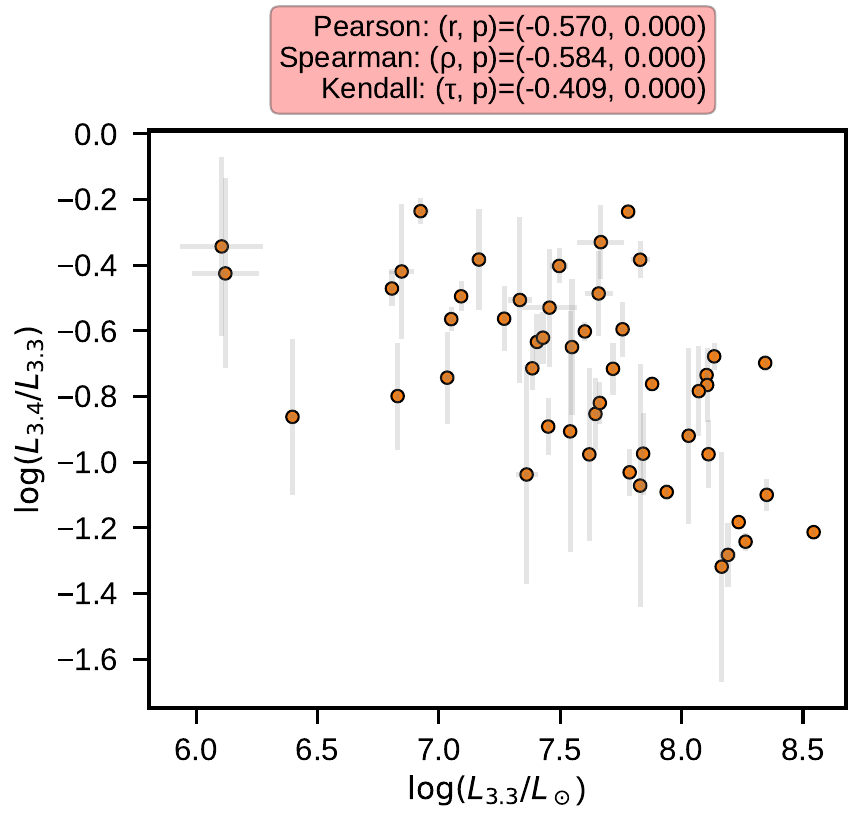}
    \includegraphics[width=0.32\hsize]{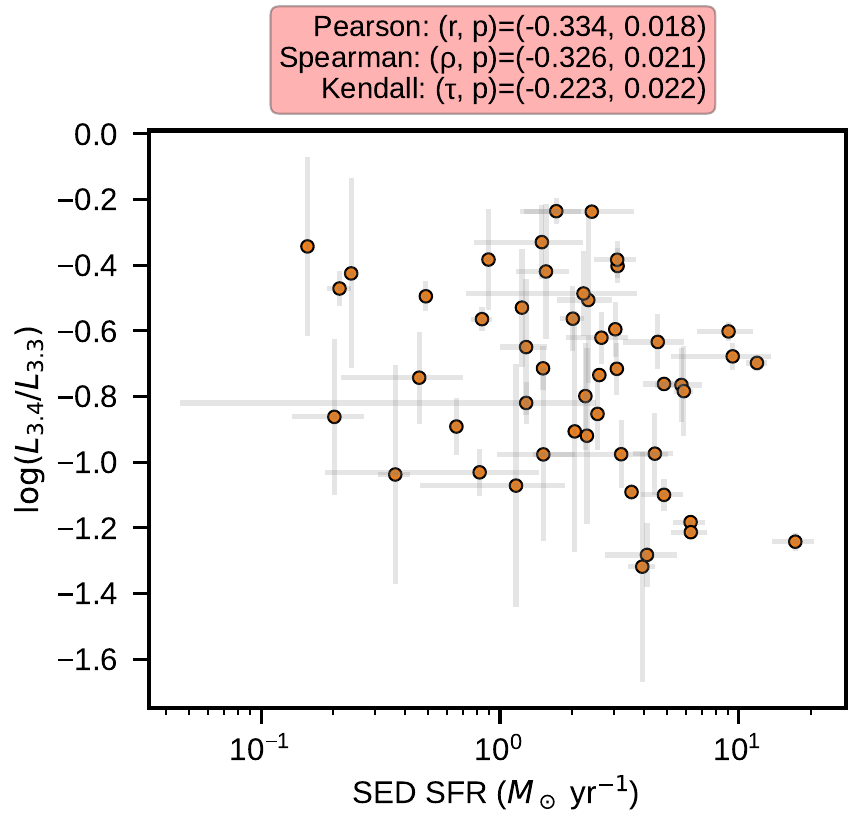}
    \caption{The aliphatic fraction of PAHs
      (measured as $L_{3.4}/L_{3.3}$) as functions
      of various galaxy properties
      (i.e., redshift, stellar mass, metallicity,
      and star formation rate measured
      from $\LIR$, $L_{3.3}$ and SED).
      We denote the correlation coefficients
      and $P$-values of Pearson, Spearman,
      Kendall in the text box. }
    \label{fig:Aliphacity_inspect}
\end{figure*}

In Figure~\ref{fig:Aliphacity_inspect}, we compare
the observed aliphatic fraction measured as
$L_{3.4}/L_{3.3}$, the ratio of the 3.3$\mum$ luminosity
to the 3.4$\mum$ luminosity, to various galaxy properties
such as redshift, stellar mass, metallicity,
and star formation rate, and conduct various
correlation tests (Pearson, Spearman, and Kendall).
We discuss the results and implications below. 

\subsubsection{PAH Aliphacity vs. Redshift}
We plot the aliphacity of PAHs (measured as $L_{3.4}/L_{3.3}$) versus redshift
in the top-left panel of Figure~\ref{fig:Aliphacity_inspect}. All the
correlation tests conclude a near zero correlation. In other word, the PAH
aliphacity in galaxies does not evolve within the observed redshift range. 

The lack of redshift dependence of the PAH aliphacity suggests that the balance
between aliphatization and aromatization of PAHs is influenced more by local
galactic conditions than by cosmic evolution. This stability spans the redshift
range of $z$\,$\simali$0.2--0.5, corresponding to a cosmic time of about 2.5 to
5 billion years ago. PAH processing timescales, including formation, evolution,
and destruction, are relatively short, often occurring over timescales of $10^6$
to $10^8\yr$, depending on environmental factors such as UV radiation intensity
and the presence of shocks. For instance, in hot gas environments, PAH lifetimes
can be as short as a few thousand years due to collisions with energetic
particles \citep{Micelotta2010}. In interstellar shocks, PAHs can be destroyed
on timescales of a few hundred million years \citep{Micelotta2010b}. These
relatively rapid processing timescales allow PAH populations to reach
equilibrium states that reflect local conditions, leading to the observed
consistency in $L_{3.4}/L_{3.3}$ across different redshifts.

\subsubsection{PAH Aliphacity vs. Stellar Mass}

Stellar mass serves as a fundamental tracer of galaxy evolution, correlating
with properties such as star formation history, dust content, and ISM
conditions, which may influence PAH spectral variations. More massive galaxies
tend to have older stellar populations, different ISM environments, and a higher
dust-to-gas ratio, all of which could impact the balance between aromatic and
aliphatic PAH emission. Additionally, given that PAHs are sensitive to local
radiation fields, variations in stellar populations across different mass
regimes might affect PAH processing, including hydrogenation and dehydrogenation
cycles. If the 3.4/3.3 feature ratio were linked to the bulk properties of a galaxy,
a dependence on stellar mass might be expected, providing insight into the role
of large-scale galactic structure in shaping PAH emission.

However, as shown in the upper middle panel in
Figure~\ref{fig:Aliphacity_inspect}, there is no significant correlation between
the 3.4/3.3 feature ratio and stellar mass, indicating that the global galaxy
properties do not play a dominant role in regulating this ratio.

\subsubsection{PAH Aliphacity vs. Metallicity}
In principle, the aliphatic fractions of PAHs
in galaxies could be related to the galaxy metallicity.
If the 3.4$\mum$ emission band indeed arises from
the aliphatic sidegroups (e.g., --CH$_3$ and --CH$_2$--)
attached to PAHs, one may expect a lower aliphatic
fraction (i.e., a lower $L_{3.4}/L_{3.3}$ ratio) in galaxies
with lower metallicities. This is because, as discussed
earlier in \S\ref{sec:L3.3_Metal}, in low-metallicity
galaxies the starlight radiation is harder and more
intense due to the reduced shielding by dust.
In such hostile environments, the aliphatic sidegroups
are more likely stripped off from the parent PAH molecules
by UV photons and therefore one expects a lower PAH
aliphacity. However, as shown in the top-right panel
of Figure~\ref{fig:Aliphacity_inspect}, we basically see
no correlation between the aliphacity and the galaxy metallicity.

Alternatively, the 3.4$\mum$ band could originate
from superhydrogenated PAHs whose edges contain
excess H atoms, i.e., some peripheral C atoms
have two H atoms. The extra H atom converts
the originally aromatic ring into an aliphatic ring
and creates two aliphatic C--H stretches that may
be responsible for the 3.4$\mum$ feature
\citep{Bernstein1996, Sandford2013,
Steglich2013, Yang2020}.
In this case, in low metallicity environments
(with a low C/H abundance), there are plenty of
H atoms available to superhydrogenate
a PAH molecule. Therefore, one would expect
a higher aliphatic fraction for PAHs
in low-metallicity galaxies.
However, we do not see in
Figures~\ref{fig:Aliphacity_inspect}
any appreciable increase of $L_{3.4}/L_{3.3}$
toward low metallicity.

On the other hand, the 3.4$\mum$ emission
feature could also be (partly) due to
the anharmonicity of the aromatic C--H stretches
\citep[see][]{Barker1987, Maltseva2016}.
In a harmonic oscillator, the spacing between all adjacent
vibrational energy levels is constant, hence the $\Delta v=1$
vibrational transitions between high $v$ levels result in
the same spectral line as that of
the $v =1\rightarrow0$ transition
(where $v$ is the vibrational quantum number).
In contrast, anharmonicity would continuously
decrease the spacing between the adjacent
vibrational states for higher values of $v$,
and therefore the $\Delta v=1$ transitions
between higher $v$ levels occur at increasingly
longer wavelengths. The anharmonicity hypothesis
interprets the weaker feature at 3.4$\mum$
as the $v=2\rightarrow1$ ``hot band''
of the 3.3$\mum$ fundamental $v =1\rightarrow0$
aromatic C--H stretching mode (see \citealt{Barker1987}).

The exact origin
of the 3.4$\mum$ band  is currently  under debate (see \citealt{Yang2017b, Tokunaga2021}). A more thorough
exploration of the PAH aliphacity and its variation
with physical and chemical conditions would
provide valuable insights into the nature of
this band. It would be interesting, for example, to examine
the 3.3 and 3.4$\mum$ emission bands
in the Small Magellanic Cloud (SMC)
where the metallicity is lower than
that of the Galactic ISM by a factor
of $\simali$5 (e.g., see \citealt{Li2002}). 

\subsubsection{ PAH Aliphacity vs. SFR}
To explore how star formation influences
the balance between the aromatization
and aliphatization of PAHs, we now examine
the relationship between $L_{3.4}/L_{3.3}$
and various star-formation indicators.

As shown in the bottom panels
in Figure~\ref{fig:Aliphacity_inspect},
there are clear negative correlations
between $L_{3.4}/L_{3.3}$ and all three
star-formation indicators
(i.e., $L_{\rm IR}$, $L_{3.3}$, and SED-derived SFR).
The $p$-values of all three tests are well below
the typical threshold of 0.05, indicating the correlations
are statistically significant.


These results suggest that the aliphatic C--H bonds
are sensitive to photodissociation in UV-rich environments,
e.g., UV photons from active star-forming regions
preferentially strip off the aliphatic sidegroups from
PAH molecules \citep{Allamandola1985, Tielens2008}.
This aligns with experimental studies showing that aliphatic
bonds are less stable and more prone to UV-induced dissociation
compared to aromatic C--H bonds
\citep[e.g.,][]{Sandford2013}. { Based on AKARI observations of the prototypical photon-dominated region (PDR) NGC 7023, \cite{Pilleri2015} showed the fraction of the 3.4 $\mum$ and 3.3 $\mu$m band intensity ($I_{3.4}/I_{3.3}$) decreases by a factor of 4 with increasing UV-photon exposure, fully consistent with the suggested scenario. For low-$z$ galaxies,} \cite{Lai2020} adopted the continuum slope between
15 and 30 $\mum$ to trace the intensity of the radiation field and reported a weak
trend of decreasing $L_{3.4}/L_{3.3}$ with the increasing of radiation field. In comparison, our
results are statistically more significant. 

The suppression of PAH aliphacity in galaxies
with intense star formation suggests that PAH
composition traces local UV radiation fields.
In active star-forming regions, lower $L_{3.4}/L_{3.3}$
reflects the preferential destruction of aliphatic bonds
by UV photons, while more quiescent environments
preserve these bonds. This highlights how PAH restructuring
is driven by UV processing rather than metallicity
or global galaxy properties. The resilience of aromatic
C--H bonds in UV-dominated regions explains
the consistent detection of the 3.3$\mum$ band
across diverse galaxy types and redshifts,
reinforcing its value as a tracer of star formation
in harsh environments
\citep[e.g.,][]{Tielens2008, Kim2012, Tacconi-Garman2013}.
The $L_{3.4}/L_{3.3}$ ratio thus offers insight into the
interplay between UV radiation, ISM chemistry,
and star formation activity over cosmic time.

\subsubsection{PAH Aliphacity vs. Galaxy Morphology}
\begin{figure}
\begin{center}
\includegraphics[width=1.0\hsize]{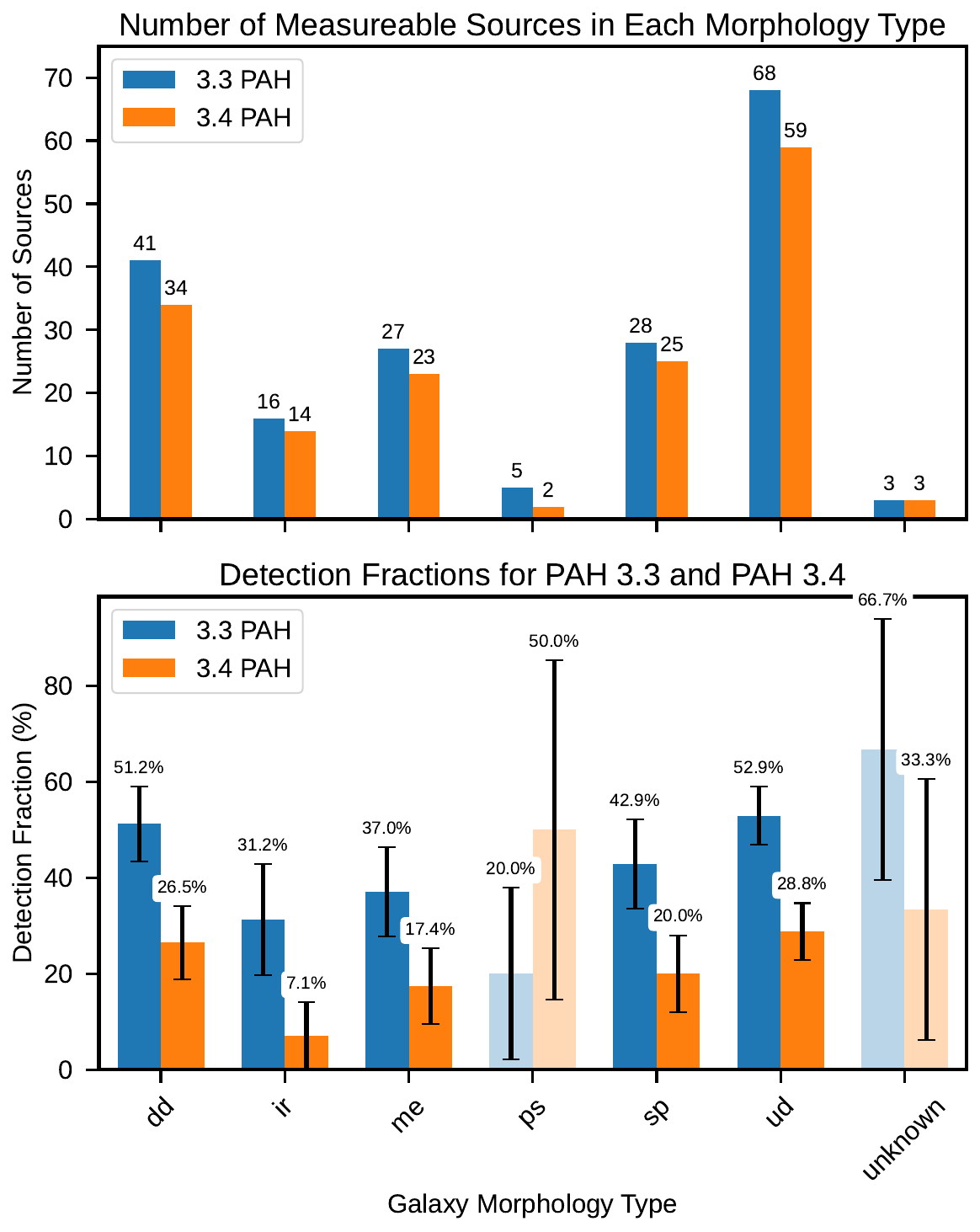}
\end{center}
\caption{Top panel: Histograms of sources
with useful NIRCam/WFSS data to determine
the 3.3 and 3.4$\mum$ emission
against galaxy morphology type.
Bottom panel: Detection rates of the 3.3 and
3.4$\mum$ emission bands among different
galaxy morphology types.
         }
         \label{fig:pah_det_morp}
\end{figure}

%
We now investigate the detection rates of
the 3.3 and 3.4$\mum$ emssion bands
in galaxies of different morphological types,
including undisturbed disks, disturbed disks,
spheroids, point sources, irregulars, and mergers.
A contingency table was constructed to compare
the detection of the 3.3 and 3.4$\mum$ features
across these six morphology categories.
We applied a chi-square test to assess
whether the detection frequencies depend on morphology.
The results, as shown in Figure~\ref{fig:pah_det_morp},
yield $p$-values of $\simali$0.167 for the 3.3$\mum$ feature
and $\simali$0.511 for the 3.4$\mum$ feature---both
exceeding the 0.05 threshold, indicating no statistically
significant difference in the feature detection rates
across different galaxy morphologies.

\begin{figure}
\begin{center}
\includegraphics[width=1.0\hsize]{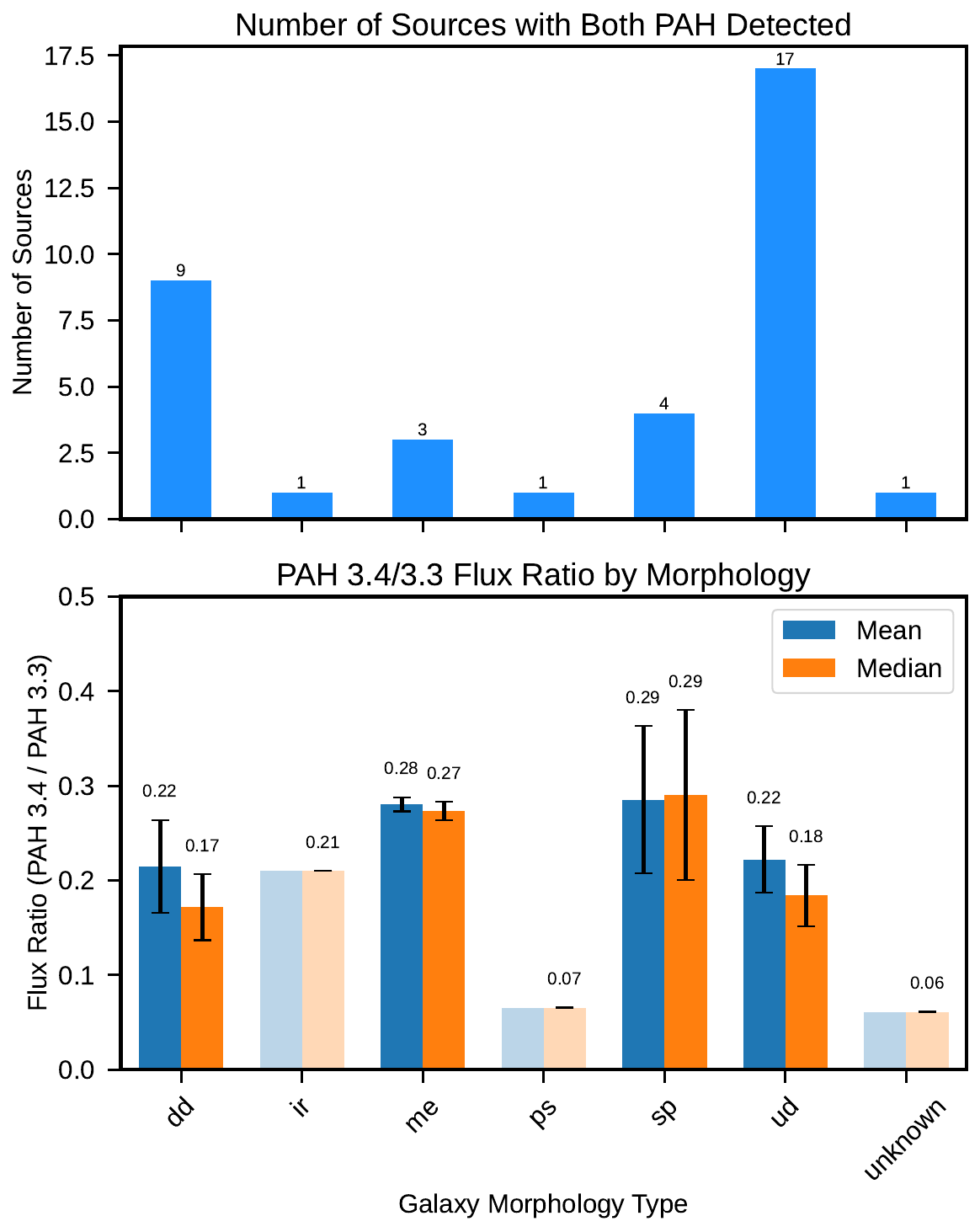}
\end{center}
\caption{Top panel: Histograms of sources
with both the 3.3 and 3.4$\mum$ bands detected
against galaxy morphology type.
Bottom panel: the mean and median values
of $L_{3.4}/L_{3.3}$ by morphology types. 
         }
         \label{fig:pah_frac_morp}
\end{figure}

To further explore potential trends,
we examine the distribution of $L_{3.4}/L_{3.3}$
across the same morphological classifications,
as shown in Figure~\ref{fig:pah_frac_morp}.
Although there is considerable scatter,
the overall distribution of $L_{3.4}/L_{3.3}$
appears similar across different morphological types,
suggesting no strong dependence on galaxy morphology.

These results suggest that the physical processes
regulating the balance between PAH aromatization
and aliphatization are largely independent of
large-scale structural features.
The lack of correlation implies that
the local star-formation activities,
rather than the galaxy global morphology,
dominate the PAH chemical structure.
This is consistent with the idea that
the UV radiation fields, driven by localized
star formation, primarily modify PAH composition,
regardless of whether the host galaxy is
a merger, disk, or irregular system.

While mergers and irregulars exhibit
a slightly broader range of $L_{3.4}/L_{3.3}$ ratios,
this likely reflects the diversity of star formation
activities within these systems rather than
morphological influences. Overall, the uniformity
in $L_{3.4}/L_{3.3}$ ratios across different morphologies
highlights the robustness of PAH emission 
as tracers of star formation and UV processing,
regardless of galaxy types.

%




\section{Summary}\label{sec:summary}
We have presented a comprehensive survey for the 3.3$\mum$ aromatic and
3.4$\mum$ aliphatic C--H stretching emission in 200 galaxies at
$z$\,$\simali$0.2--0.5 using JWST/NIRCam WFSS data from the JWST Cycle~1 legacy
program FRESCO. This study extends the detection and characterization of these
PAH features well beyond the local Universe, probing galaxies with IR
luminosities as low as $L_{\rm IR}\sim 10^9$--$10^{10}\Lsun$, significantly
lower than in previous extragalactic PAH studies from e.g., AKARI or {\it
Spitzer}. Our major results are:
\begin{enumerate}
\item In terms of detections, the 3.3$\mum$
aromatic feature is detected in 88 of 187 galaxies
and the 3.4$\mum$ aliphatic feature is detected
in 37 of 159 galaxies. Sources without detections
are mostly due to their weak PAH emission which
fails below the NIRCam/WFSS detection limit;
\item The 3.3$\mum$ aromatic emission ($L_{3.3}$)
is strongly correlated with the galaxy IR luminosity
($\LIR$) over three orders of magnitude.
Meanwhile, the relative strength of this feature
to the galaxy IR lumionosity ($L_{3.3}/\LIR$)
shows a strong dependence on galaxy metallicity,
with lower metallicity galaxies exhibiting weaker PAH
emission;
\item The 3.3$\mum$ aromatic emission strength is also closely correlated with
the SFRs derived either from the galaxy IR luminosity or from the galaxy SED. We
have demonstrated that the 3.3$\mum$ PAH emission is a valid SFR indicator (see
Equation \,\ref{equ:sed-pah}) for galaxies with a broad
range of SFRs from $\simali$0.1$\Msun\yr^{-1}$ to
$\simali$300$\Msun\yr^{-1}$ with an intrinsic scatter of $\simali$0.2--0.3 dex. This tight correlation establishes the usage of the 3.3$\mum$ PAH emission
to infer SFRs in large statistical samples of galaxies across cosmic time with
JWST and future IR telescopes;
\item The ratios of the 3.4 and 3.3$\mum$ emission
($L_{3.4}/L_{3.3}$) show a range of variations from $\simali$0.05 to $\simali$0.58, with a median value at $\simali$0.19, indicating PAH aliphatic fractions (measured as the fractions of carbon atoms in aliphatic units) of $\simali$0.78\%--8.3\% with a median value of $\simali$2.9\%.
A negative correlation between $L_{3.4}/L_{3.3}$
and the SFR indicators (e.g., $L_{\rm IR}$, $L_{3.3}$, and SED-derived SFR) suggests that UV photons from active star-forming regions strip off the alipatic sidegroups from PAH molecules;
\item We do not detect any significant evolution of
PAH aliphatic fractions (as measured by $L_{3.4}/L_{3.3}$)
with galaxy redshift, stellar mass, metallicity, or morphology,
indicating that the aliphatic component of PAHs
is more sensitive to local star formation conditions
than to global galaxy properties. 
\end{enumerate}

Our work demonstrates the power of the JWST NIRCam/WFSS mode in surveying PAH
molecules in statistically significant, less-biased samples of galaxies,
providing a more comprehensive view of PAH properties and star formation
processes across cosmic time. This study paves the way for future programs to
explore the nature, origin and evolution of PAHs and their role in galaxy
formation beyond the local Universe.

\begin{acknowledgements}

JL is supported in part by JWST Mid-Infrared Instrument (MIRI) grant No.
80NSSC18K0555, and the NIRCam science support contract NAS5-02105, both from
NASA Goddard Space Flight Center to the University of Arizona; AL is supported
in part by STScI Award JWST-AR-03879.001-A. XJY is supported in part by NSFC\,12333005 and 12122302.

This work is based on observations made with the  NASA/ESA/CSA James Webb Space
Telescope. The data were obtained from the Mikulski Archive for Space Telescopes
at the Space Telescope Science Institute, which is operated by the Association
of Universities for Research in Astronomy, Inc., under NASA contract NAS 5-03127
for JWST. { These observations are associated with program 1895 (FRESCO, PI: Pascal Oesch), which can be accessed via \dataset[doi:10.17909/gdyc-7g80]{https://doi.org/10.17909/gdyc-7g80}}. The
authors acknowledge the FRESCO team for developing their observing program with
a zero-exclusive-access period.

\end{acknowledgements}

\facilities{JWST (NIRCam)}

\software{Astropy \citep{Astropy}, Matplotlib \citep{Hunter2007}, NumPy \citep{Harris2020}, SciPy \citep{Virtanen2020}}

\bibliography{main_bibliography}{}
\bibliographystyle{aasjournal}

\end{document}